{}
{}

\documentclass[10pt,a4paper]{article}

\newif\ifpublic\publictrue

\ifpublic\else\usepackage{showkeys}\fi
\usepackage[bookmarks=true,hyperfigures=true,hidelinks,bookmarksnumbered,unicode]{hyperref}
\usepackage[nosort]{cite}
\usepackage[parsep]{collref}
\usepackage[english]{babel}
\usepackage[T1]{fontenc}
\usepackage[utf8]{inputenc}
\usepackage{lmodern}
\usepackage{amsfonts,amsbsy,dsfont,euscript,mathrsfs,fixmath}
\usepackage{amsmath,amssymb}
\usepackage{scalerel}
\usepackage{enumerate}
\usepackage{graphicx}

\def\showkeysrefformat#1{{\normalfont\tiny\ttfamily#1}}
\makeatletter
\def\SK@@ref#1>#2\SK@{{\@inlabelfalse\leavevmode\vbox to\z@{\vss\SK@refcolor\rlap{\vrule\raise .75em \hbox{\showkeysrefformat{#2}}}}}}
\makeatother

\usepackage[a4paper,text={160mm,247mm},centering]{geometry}
\allowdisplaybreaks[3]
\numberwithin{equation}{section}
\def\[{\begin{equation}\begin{aligned}}
\def\]{\end{aligned}\end{equation}}

\usepackage[font=small,labelfont=bf,width=0.85\textwidth]{caption}

\expandafter\def\expandafter\bfseries\expandafter{\bfseries\ifmmode\else\boldmath\fi}
\expandafter\def\expandafter\mdseries\expandafter{\mdseries\ifmmode\else\unboldmath\fi}
\expandafter\def\expandafter\normalfont\expandafter{\normalfont\ifmmode\else\unboldmath\fi}

\RequirePackage{verbatim}

\makeatletter
\newwrite\bibinl@out
\newenvironment{bibtex}[1][\jobname]{%
\immediate\openout\bibinl@out #1.bib%
\immediate\write\bibinl@out{\@percentchar generated from `\jobname' starting line \the\inputlineno^^J}%
\def\verbatim@processline{\immediate\write\bibinl@out{\the\verbatim@line}}%
\@bsphack\let\do\@makeother\dospecials\catcode`\^^M\active\verbatim@start%
}
{\immediate\closeout\bibinl@out\@esphack}
\makeatother

\let\barefrac=\frac
\renewcommand{\frac}[2]{\mathinner{\barefrac{#1}{#2}}}

\let\baresqrt=\sqrt
\makeatletter
\renewcommand{\sqrt}{\@ifnextchar[\@sqrt@space@a\@sqrt@space@b}
\def\@sqrt@space@a[#1]#2{\mathinner{\mathchoice{\mkern-3mu}{\mkern-3mu}{}{}\baresqrt[#1]{#2}}}
\def\@sqrt@space@b#1{\mathinner{\mathchoice{\mkern-3mu}{\mkern-3mu}{}{}\baresqrt{#1}}}
\makeatother

\makeatletter
\let\per@dot@old=\.
\def\.{\ifmmode\def\per@dot@sel{\mkern3mu}\else\def\per@dot@sel{\per@dot@old}\fi\per@dot@sel}
\makeatother

\let\barefootnote=\footnote
\renewcommand{\footnote}[1]{\barefootnote{#1\vspace{3pt}}}

\newcommand{\vfrac}[2]{\ifmmode\mathinner{\textstyle^{#1}\!/\!_{#2}}\else$^{#1}\!/\!_{#2}$\fi}

\DeclareMathOperator{\Tr}{Tr}

\newcommand{\Real}{\mathds{R}}

\newcommand{\Integer}{\mathds{Z}}

\newcommand{\ind}[1]{{\scriptscriptstyle{#1}}}

\makeatletter
\newcommand*\bigcdot{\mathpalette\bigcdot@{.5}}
\newcommand*\bigcdot@[2]{\mathbin{\vcenter{\hbox{\scalebox{#2}{$\m@th#1\bullet$}}}}}
\makeatother

\newcommand{\al}{\alpha}
\newcommand{\be}{\beta}
\newcommand{\ga}{\gamma}
\newcommand{\tg}{\tilde{\gamma}}

\newcommand{\om}{\omega^b}
\newcommand{\tom}{\omega^a}

\usepackage{mathtools}

\newcommand{\alg}[1]{\mathfrak{#1}}
\newcommand{\grp}[1]{\mathrm{#1}}

\DeclareMathOperator{\Lie}{Lie}

\newcommand{\com}[2]{[#1,#2]}

\def\<{\big\langle}
\def\>{\big\rangle}

\newcommand{\geom}[1]{\mathrm{#1}}

\newcommand{\AdS}{\geom{AdS}}

\newcommand{\Sp}{\geom{S}}

\newcommand{\To}{\geom{T}}

\newcommand{\Lag}{\mathcal{L}}
\newcommand{\Act}{\mathcal{S}}
\newcommand{\Ham}{\mathcal{H}}
\DeclareFontEncoding{LS1}{}{}
\DeclareFontSubstitution{LS1}{stix}{m}{n}
\DeclareSymbolFont{stixsymbols}{LS1}{stixscr}{m}{n}
\SetSymbolFont{stixsymbols}{bold}{LS1}{stixscr}{b}{n}
\DeclareMathSymbol{\kay}{\mathalpha}{stixsymbols}{"6B}
\DeclareMathSymbol{\hay}{\mathalpha}{stixsymbols}{"68}
\DeclareMathAlphabet{\mathdsl}{U}{bbm}{m}{sl}
\newcommand{\Tstr}{T_{\mathrm{str}}}
\newcommand{\field}{\Psi}
\newcommand{\fields}{\Psi}

\newcommand{\ket}[1]{\big| #1 \big \rangle }

\providecommand{\href}[2]{#2}

\makeatletter
\def\mr@ignsp#1 {\ifx\:#1\@empty\else #1\expandafter\mr@ignsp\fi}
\newcommand{\multiref}[1]{\begingroup%
\xdef\mr@no@sparg{\expandafter\mr@ignsp#1 \: }%
\def\mr@comma{}\def\mr@dash{-}%
\@for\mr@refs:=\mr@no@sparg\do{%
\ifx\mr@refs\mr@dash\def\mr@comma{}--\else%
\mr@comma\def\mr@comma{,}\ref{\mr@refs}\fi}%
\endgroup}
\renewcommand{\eqref}[1]{(\multiref{#1})}
\makeatother

\makeatletter
\newcommand{\namedref}[2]{\hyperref[#2]{#1~\ref*{#2}}}
\newcommand{\secref}{\@ifstar{\namedref{Section}}{\namedref{sec.}}}
\newcommand{\appref}{\@ifstar{\namedref{Appendix}}{\namedref{app.}}}
\newcommand{\tabref}{\@ifstar{\namedref{Table}}{\namedref{tab.}}}
\newcommand{\figref}{\@ifstar{\namedref{Figure}}{\namedref{fig.}}}
\makeatother

\let\oldbib=\thebibliography
\def\thebibliography{\phantomsection\addcontentsline{toc}{section}{\refname}\oldbib}

\let\oldtoc=\tableofcontents
\def\tableofcontents{\phantomsection\addcontentsline{toc}{section}{\contentsname}\oldtoc}

\providecommand{\hypersetup}[1]{}
\providecommand{\texorpdfstring}[2]{#1}
\hypersetup{plainpages=false}
\hypersetup{pdfpagemode=UseNone}
\hypersetup{bookmarksnumbered=true}
\hypersetup{pdfstartview=FitH}
\hypersetup{colorlinks=false}
\hypersetup{citebordercolor={1 1 1}}
\hypersetup{urlbordercolor={1 1 1}}
\hypersetup{linkbordercolor={1 1 1}}

\makeatletter
\let\@keywords\@empty
\let\@subject\@empty
\providecommand{\keywords}[1]{\gdef\@keywords{#1}}
\providecommand{\subject}[1]{\gdef\@subject{#1}}
\def\thetitle{\@title}
\def\theauthor{\@author}
\def\thesubject{\@subject}
\def\thedate{\@date}
\def\thekeywords{\@keywords}
\makeatother
\AtBeginDocument{
\hypersetup{pdftitle={\thetitle}}
\hypersetup{pdfauthor={\theauthor}}
\hypersetup{pdfsubject={\thesubject}}
\hypersetup{pdfkeywords={\thekeywords}}}

\newif\ifshownote
\shownotetrue

\ifpublic\shownotefalse\fi

\ifshownote

\ifpdf\else\RequirePackage[active]{srcltx}\fi
\RequirePackage{xcolor}
\RequirePackage{ulem}

\newcommand{\remark}[2][]{{\normalfont\sffamily\hspace{1ex}
\def\emph{\textsl}\def\textbullet{$\bullet$}
\def\tmparga{#1}
\def\tmpargb{BH}\ifx\tmparga\tmpargb\color[rgb]{0.7,0,0}\fi
\def\tmpargb{AR}\ifx\tmparga\tmpargb\color[rgb]{0,0.7,0}\fi
\def\tmpargb{FS}\ifx\tmparga\tmpargb\color[rgb]{0,0,0.7}\fi
\def\tmpargb{}\ifx\tmparga\tmpargb\normalfont\color{red}\fi
\def\tmpargb{}\ifx\tmparga\tmpargb\else \textbf{#1:}\fi
#2\hspace{1ex}}}

\else

\newcommand{\remark}[2][]{\ignorespaces}

\fi

\title{Elliptic deformations of the \texorpdfstring{$\AdS_3 \times \Sp^3 \times \To^4$}{AdS3 × S3 × T4} string}
\author{Ben Hoare\texorpdfstring{$^{a,}$\footnote{\texttt{ben.hoare@durham.ac.uk}}}{}, Ana L.~Retore\texorpdfstring{$^{a,}$\footnote{\texttt{ana.retore@durham.ac.uk}}}{} and Fiona K.~Seibold\texorpdfstring{$^{b,}$\footnote{\texttt{f.seibold21@imperial.ac.uk}}}{}}

\begin{document}

\pdfbookmark[1]{Title Page}{title}
\thispagestyle{empty}

\vspace*{2cm}
\begin{center}
\begingroup\Large\bfseries\thetitle\par\endgroup
\vspace{1cm}

\renewcommand{\thefootnote}{\roman{footnote}}
\begingroup\theauthor\par\endgroup
\vspace{1cm}

\textit{$^a $Department of Mathematical Sciences, Durham University, Durham DH1 3LE, UK}\\
\vspace{1mm}
\textit{$^b $Blackett Laboratory, Imperial College, London SW7 2AZ, UK}
\vspace{5mm}

\vfill

\textbf{Abstract}\vspace{5mm}

\begin{minipage}{12.5cm}
With the aim of investigating the existence of an integrable elliptic deformation of strings on $\AdS_3 \times \Sp^3 \times \To^4$, we compute the tree-level worldsheet S-matrix of the elliptically-deformed bosonic sigma model on $\AdS_3 \times \Sp^3$ in uniform light-cone gauge.
The resulting tree-level S-matrix is compatible with the integrability of the model and has interesting features, including a hidden $\grp{U}(1)$ symmetry not manifest in the Lagrangian.
We find that it cannot be embedded in the known exact integrable R-matrices describing deformations of the undeformed $\AdS_3 \times \Sp^3 \times \To^4$ light-cone gauge S-matrix including fermions.
Therefore, we construct embeddings of the deformed 6-d metric in type II supergravity with constant dilaton and homogeneous fluxes.
The simplicity of these solutions suggests they are promising candidates to lead to an integrable string sigma model including fermions.
\end{minipage}

\vspace*{4cm}

\end{center}

\setcounter{footnote}{0}
\renewcommand{\thefootnote}{\arabic{footnote}}

\newpage

\section{Introduction}\label{sec:introduction}

The construction of the Yang-Baxter (YB) deformation~\cite{Delduc:2013qra,Kawaguchi:2014qwa,Klimcik:2002zj} of the Metsaev-Tseytlin supercoset sigma model for semi-symmetric spaces \cite{Metsaev:1998it,Berkovits:1999zq} has led to substantial interest in integrable deformations of strings on $\AdS$ backgrounds.
YB deformations can be grouped into two classes: homogeneous and inhomogeneous, depending on whether the deforming operator satisfies the classical YB equation or its modified counterpart.
On the other hand, the $\grp{SL}(2;\Real) \times \grp{SU}(2)$ principal chiral model, the sigma model with target space $\AdS_3 \times \Sp^3$, has long been known to admit a hierarchy of integrable deformations -- rational (undeformed), trigonometric and elliptic~\cite{Cherednik:1981df}.
\unskip\footnote{Recently, it has also been shown that it is possible to generalise and construct an elliptic deformation of the principal chiral model for higher rank groups~\cite{Lacroix:2023qlz}.}
The adjectives refer to the periodicity of the spectral parameter in the Lax connection.
For a recent review on integrable deformations of sigma models, see~\cite{Hoare:2021dix}.

The Metsaev-Tseytlin supercoset sigma model for $\Integer_4$ permutation supercosets, a special class of semi-symmetric spaces whose superisometry group is of product form, can be used to describe strings on $\AdS_3$ backgrounds supported by Ramond-Ramond (R-R) fluxes in the Green-Schwarz formulation~\cite{Babichenko:2009dk}.
\unskip\footnote{More precisely, the supercoset sigma model describes the sector of the theory associated with the curved part of the geometry, i.e.~$\AdS_3 \times \Sp^3$ in the case of the $\AdS_3 \times \Sp^3 \times \To^4$.
To embed the supercoset in the Green-Schwarz sigma model additional bosonic fields need to be included to describe the $\To^4$, along with fermionic fields ensuring $\kappa$-symmetry.
The integrability of the supercoset sigma model is expected to extend to the Green-Schwarz sigma model as shown to second order in fermions in~\cite{Wulff:2014kja}.}
Starting from the inhomogeneous Drinfel'd-Jimbo bi-YB deformation~\cite{Hoare:2014oua,Klimcik:2008eq} of the supercoset sigma model for the semi-symmetric space
\begin{equation}
\frac{\grp{PSU}(1,1|2) \times \grp{PSU}(1,1|2)}{\grp{SU}(1,1) \times \grp{SU}(2)}~,
\end{equation}
an integrable trigonometric deformation of strings on $\AdS_3 \times \Sp^3 \times \To^4$ was constructed in~\cite{Seibold:2019dvf,Hoare:2022asa}.
Motivated by these results, our goal in this paper is to investigate the existence of an integrable elliptic deformation of strings on $\AdS_3 \times \Sp^3 \times \To^4$.

\medskip

In the space of integrable string sigma models and their deformations, the $\AdS_3 \times \Sp^3 \times \To^4$ background has special properties stemming from the product structure of its superisometry group $\grp{PSU}(1,1|2) \times \grp{PSU}(1,1|2)$.
\unskip\footnote{The same is true of the closely related $\AdS_3 \times \Sp^3 \times \Sp^3 \times \Sp^1$ background whose superisometry group is $\grp{D}(2,1;\alpha) \times \grp{D}(2,1;\alpha)$.}
In particular, when only one factor is deformed, which we refer to as a unilateral deformation, then at least half the supersymmetry will be preserved.
Moreover, for the unilateral inhomogeneous YB deformation constructed from the Drinfel'd-Jimbo R-matrix, the resulting geometry is smooth with vanishing Ricci scalar and the dilaton is constant~\cite{Seibold:2019dvf,Hoare:2022asa}.

The unilateral inhomogeneous Drinfel'd-Jimbo YB deformation of the $\AdS_3 \times \Sp^3$ supercoset sigma model is closely related to the trigonometric deformation of $\AdS_3 \times \Sp^3$~\cite{Klimcik:2002zj}.
The deformed bosonic sigma models are the same up to a total derivative.
While the YB deformation leads a particular choice of supporting R-R fluxes, by applying sequences of TsT transformations in the 4-torus directions it is possible to write the background in an $\grp{SO}(4)$-invariant way~\cite{Hoare:2022asa}, where $\grp{SO}(4)$ is locally isomorphic to the maximal compact subgroup of the T-duality group of 6-d maximal supergravity.
\unskip\footnote{This can be extended to include the Neveu-Schwarz-Neveu-Schwarz (NS-NS) 3-form flux starting from the unilateral inhomogeneous Drinfel'd-Jimbo bi-YB deformation~\cite{Delduc:2018xug} of the $\AdS_3 \times \Sp^3$ mixed flux supercoset sigma model~\cite{Cagnazzo:2012se}.
Again by applying sequences of TsT transformations, the background can be written in an $\grp{SO}(5)$-invariant way~\cite{Hoare:2022asa}, where $\grp{SO}(5)$ is locally isomorphic to the maximal compact subgroup of the U-duality group of 6-d maximal supergravity.}
We refer to this model as the trigonometric deformation of strings on $\AdS_3 \times \Sp^3 \times \To^4$.

The properties of the trigonometric deformation motivate us to investigate the existence of an analogous elliptic deformation.
Since an integrable elliptic deformation of the Metsaev-Tseytlin supercoset sigma model is not known, our main approach will be to consider the elliptically-deformed bosonic sigma model on $\AdS_3 \times \Sp^3$ and investigate the tree-level worldsheet S-matrix in uniform light-cone gauge.

\medskip

In order to quantise the worldsheet sigma model in the Green-Schwarz formalism, one approach is to fix a uniform light-cone gauge isolating the physical degrees of freedom~\cite{Arutyunov:2005hd,Arutyunov:2009ga,Arutyunov:2014jfa}.
This relies on a choice of two commuting isometries, with different choices leading to different gauge-fixed theories related by $T\bar{T}$, $J\bar{T}$ transformations and their relatives~\cite{Frolov:2019xzi,Borsato:2023oru}.
Whatever choice is made, the same string spectrum should eventually be the same.
Typically, in the study of strings on $\AdS_3 \times \Sp^3 \times \To^4$, the diagonal isometries $\grp{U}(1)_{\textrm{diag}} \subset \grp{SL}(2,\Real) \times \grp{SL}(2,\Real)$ and $\grp{U}(1)_{\textrm{diag}} \subset \grp{SU}(2) \times \grp{SU}(2)$ are used to fix light-cone gauge, see, e.g., \cite{Borsato:2013qpa,Sundin:2013ypa,Borsato:2014exa,Sundin:2014ema} and \cite{Demulder:2023bux} for a review.
However, these do not correspond to symmetries of the elliptically-deformed $\AdS_3 \times \Sp^3$ sigma model, and we are forced to use different isometries.
As a result, the undeformed limit of our results differ from those in the literature by terms that can be interpreted in terms of $J\bar{T}$ deformations.
Since our main interest is in the elliptic deformation, other than observing the relation between the light-cone gauge-fixed models, we do not explore this further.
For a detailed study of the different choices of light-cone gauge for $\AdS_n \times \Sp^n$ string sigma models and their deformations, see~\cite{Borsato:2023oru}.

After fixing light-cone gauge, the standard route to quantisation starts by decompactifying and computing the scattering matrix for the transverse excitations on the worldsheet.
When the model is integrable, we can attempt to use symmetries and the quantum YB equation to bootstrap an exact result, up to phase factors, valid to all orders in the inverse string tension.
To determine the phase factors additional physical constraints, such as unitarity, crossing symmetry and analyticity properties need to be imposed, which has successfully been accomplished for the $\AdS_3 \times \Sp^3 \times \To^4$ superstring~\cite{Frolov:2021fmj,Frolov:2021zyc}.
The bosonic worldsheet spectrum consists of 4 gapped and 4 gapless modes, originating from $\AdS_3 \times \Sp^3$ and $\To^4$ respectively.
Supersymmetry requires that there are similarly 4 gapped and 4 gapless fermionic modes.
The full S-matrix, which is a $256 \times 256$ matrix, can be broken down into $4 \times 4$ building blocks, each describing the scattering of 1 boson and 1 fermion.

Deformations of these $4 \times 4$ blocks compatible with integrability have been classified in~\cite{deLeeuw:2020ahe,deLeeuw:2021ufg}.
In particular, there are two families known as the 6vB and 8vB R-matrices, which are of 6 vertex and 8 vertex type respectively.
The former includes the building blocks of the trigonometric deformation~\cite{Hoare:2014oua}, while the latter is naturally expressed in terms of elliptic functions.
Therefore, the existence of the 8vB R-matrix provides an additional motivation to investigate elliptic deformations of strings on $\AdS_3 \times \Sp^3 \times \To^4$.

\medskip

In this paper, we compute the tree-level worldsheet S-matrix for the elliptic deformation of the bosonic sigma model on $\AdS_3 \times \Sp^3$.
While we do not find agreement with either the 6vB or 8vB R-matrices of~\cite{deLeeuw:2020ahe,deLeeuw:2021ufg},
\unskip\footnote{Let us note already that we do not necessarily expect to match with the 8vB R-matrix of~\cite{deLeeuw:2020ahe,deLeeuw:2021ufg} that motivated us to consider this model.
Indeed, the 6vB and 8vB solutions to the quantum YB equation only coincide at the undeformed $\AdS_3 \times \Sp^3 \times \To^4$ point, while we expect the trigonometric deformation to be a special case of the elliptic deformation, assuming an integrable version of the latter exists.}
we do find a number of interesting features, including a hidden $\grp{U}(1)$ symmetry of the tree-level S-matrix, which is not a global symmetry of the light-cone gauge Lagrangian.
We discuss different options for how our result could still be extended to include fermions and remain compatible with integrability, or break integrability mildly by a non-integrable twist.

In order to test these options, and for the light-cone gauge theory to describe a string sigma model, we require an embedding of the deformed 6-d metric in type II supergravity.
With this in mind, we construct candidate embeddings with constant dilaton and homogeneous fluxes, motivated by the backgrounds in~\cite{Hoare:2022asa}.
While we leave the question of whether these backgrounds preserve any supersymmetry or lead to an integrable string sigma model for the future, the tree-level scattering amplitudes of the gapped bosonic modes associated to the deformed $\AdS_3 \times \Sp^3$ will be unchanged by the fermionic and gapless modes, hence provide a probe of the string sigma model.

The outline of this paper is as follows.
In~\secref{sec:metricAdS3xS3} we recall the elliptic deformation of the bosonic $\AdS_3 \times \Sp^3$ sigma model and introduce our parametrisation.
In \secref{sec:lcgft} we light-cone gauge fix the theory.
A discussion of the symmetries before and after light-cone gauge fixing is given in \appref{app:symmetries}.
The presence of the deformation requires us to consider an alternative light-cone gauge fixing to the one typically used in the literature, and we comment on the relation between the gauge fixings in \secref{sec:jttb}.
The tree-level S-matrix is presented in \secref{sec:tlsmat} and its relation to known integrable R-matrices is discussed in \secref{sec:exact}.
Finally, in \secref{sec:sugra} we present an embedding of the elliptic deformation of $\AdS_3 \times \Sp^3$ in type II supergravity, and conclude in \secref{sec:conc} with a summary of our results and future directions.

\section{Elliptic deformation of \texorpdfstring{$\AdS_3 \times \Sp^3$}{AdS3 × S3}}\label{sec:metricAdS3xS3}

In this section we define the action and the metric of the elliptically-deformed $\AdS_3 \times \Sp^3$ sigma model.

\subsection{Action and metric}

Both $\AdS_3$ and $\Sp^3$ are symmetric spaces, described by the cosets
\begin{equation}
\AdS_3 = \frac{\grp{SO}(2,2)}{\grp{SO}(1,2)} \cong \frac{\grp{SL}(2; \Real) \times \grp{SL}(2; \Real)}{\grp{SL}(2; \Real)}~, \qquad \Sp^3 = \frac{\grp{SO}(4)}{\grp{SO}(3)} \cong \frac{\grp{SU}(2) \times \grp{SU}(2)}{\grp{SU}(2)}~.
\end{equation}
Bosonic strings on $\AdS_3 \times \Sp^3$ can then be described in two equivalent ways: either by the symmetric space sigma model with group element $g \in \grp{SL}(2; \Real) \times \grp{SL}(2; \Real) \times \grp{SU}(2) \times \grp{SU}(2)$ and local gauge symmetry $g \rightarrow g h$ with $h \in \grp{SL}(2; \Real) \times \grp{SU}(2)$, or more simply by the principal chiral model (PCM) on $\grp{G} = \grp{SL}(2; \Real) \times \grp{SU}(2)$, without any gauge symmetry.
Here we will adopt the latter description, which is particularly suited to introduce deformation parameters into the action.
We consider the elliptically-deformed sigma model action
\begin{equation} \label{eq:Lag-sigma-model}
\Act = \frac{1}{4} \int d\tau d\sigma (\gamma^{\alpha \beta} + \epsilon^{\alpha \beta}) \Tr \left[ g^{-1} \partial_\alpha g \, \mathcal O \, g^{-1} \partial_\beta g \right]~, \qquad g \in \grp{G} = \grp{SL}(2;\Real) \times \grp{SU(2)}~,
\end{equation}
where the indices $\alpha, \beta = 0,1$ label the time and space coordinates $\sigma^0 \equiv \tau$ and $\sigma^1 \equiv \sigma$ on the 2-d string worldsheet, $\gamma^{\alpha \beta} = \sqrt{-h} h^{\alpha \beta}$ is the Weyl-invariant auxiliary metric on the worldsheet and $\epsilon^{\alpha \beta}$ is the anti-symmetric Levi-Civita symbol with $\epsilon^{01}=- \epsilon^{10}=1$.
The deforming linear operator $\mathcal O : \alg{g} \rightarrow \alg{g}$ acts on the generators of $\alg{g} = \alg{sl}(2;\Real) \oplus \alg{su}(2) = \Lie(\grp{G})$ diagonally as
\begin{equation} \label{eq:Odef}
\mathcal O(L_j) = -\alpha_j L_j~, \qquad \mathcal O (J_j) = \beta_j J_j~, \qquad j=1,2,3~.
\end{equation}
The overall coupling, which corresponds to the string tension in the context of string theory, has been absorbed into the real deformation parameters $\alpha_j$ and $\beta_j$.
The sign convention in \eqref{eq:Odef} is such that $\alpha_j=\beta_j=1$ corresponds to the undeformed PCM.
The minus sign in the operator acting on the generators of $\alg{sl}(2;\Real)$ is associated to the non-compactness of the Lie algebra.
Alternatively, we could have defined the action \eqref{eq:Lag-sigma-model} with the supertrace rather than the trace, in which case $\mathcal O$ would simply reduce to the identity in the undeformed case, allowing the action \eqref{eq:Lag-sigma-model} to be generalised to supergroups.
Our choice for the generators of $\alg{sl}(2; \Real)$ is
\unskip\footnote{The Pauli matrices are given by
\begin{equation*}
\sigma_1 = \begin{pmatrix} 0 & 1 \\ 1 & 0 \end{pmatrix}~, \qquad
\sigma_2 = \begin{pmatrix} 0 & -i \\ i & 0 \end{pmatrix}~, \qquad
\sigma_3 = \begin{pmatrix} 1 & 0 \\ 0 & -1 \end{pmatrix}~.
\end{equation*}}
\begin{equation}\label{eq:lmat}
L_1 = \begin{pmatrix} \sigma_1 & 0 \\ 0 & 0 \end{pmatrix}~, \qquad
L_2 = \begin{pmatrix} i\sigma_2 & 0 \\ 0 & 0 \end{pmatrix}~, \qquad
L_3 = \begin{pmatrix} \sigma_3 & 0 \\ 0 & 0 \end{pmatrix}~,
\end{equation}
while for $\alg{su}(2)$ we choose
\begin{equation}\label{eq:jmat}
J_1 = \begin{pmatrix} 0 & 0 \\ 0 & i\sigma_1 \end{pmatrix}~, \qquad
J_2 = \begin{pmatrix} 0 & 0 \\ 0 & i\sigma_2 \end{pmatrix}~, \qquad
J_3 = \begin{pmatrix} 0 & 0 \\ 0 & i \sigma_3 \end{pmatrix}~.
\end{equation}
To obtain the metric of the model we consider a specific parameterisation of the group-valued field $g$,
\begin{equation} \label{eq:param}
g=e^{T L_2}e^{U L_3}e^{V L_1}e^{\Phi J_2}e^{X J_3}e^{Y J_1}~.
\end{equation}
Gathering all the target space fields $\field^\mu = (T, U, V, \Phi, X, Y)^\mu$, the sigma model action \eqref{eq:Lag-sigma-model} takes the general form
\begin{equation}
\Act = -\frac{1}{2} \int d\tau d \sigma \left(\gamma^{\alpha \beta} G_{\mu \nu} \partial_\alpha \field^\mu \partial_\beta \field^\nu + \epsilon^{\alpha \beta} B_{\mu \nu} \partial_\alpha \field^\mu \partial_\beta \field^\nu\right)~,
\end{equation}
where the symmetric tensor $G_{\mu \nu}$ denotes the metric on target space and the anti-symmetric 2-form $B_{\mu \nu}$ is the B-field.
It turns out that for the elliptically-deformed $\AdS_3 \times \Sp^3$ background the B-field vanishes $B_{\mu \nu}=0$, while the metric $ds^2 = G_{\mu \nu} d \field^\mu d \field^\nu$ is given by
\begin{align}\label{eq:ellipticmetricAdS3xS3}
ds^2 &= ds^2_a + ds^2_b~, \\
ds_a^2&=\left(\al_1\sinh^22U+\cosh^22U\left(-\al_2\cosh^22V+\al_3\sinh^22V\right) \right)dT^2+\al_1dV^2+\nonumber\\
&\qquad +\left(\al_3\cosh^22V-\al_2\sinh^22V\right)dU^2 -2\al_1\sinh 2UdT dV+(\al_3-\al_2)\cosh2U\sinh4VdT dU~,\nonumber\\
ds^2_b &=\left(\be_1\sin^22X+\cos^22X\left(\be_2\cos^22Y+\be_3\sin^22Y\right)\right)d\Phi^2+\be_1dY^2+\nonumber\\
&\qquad+\left(\be_3\cos^22Y+\be_2\sin^22Y\right)dX^2 -2\be_1\sin 2Xd\Phi dY+(\be_3-\be_2)\cos2X\sin4Yd\Phi dX~.\nonumber
\end{align}
The metric decomposes into $ds^2_a$ and $ds^2_b$, the metrics of elliptically-deformed $\AdS_3$ and $\Sp^3$ respectively.
They are related by the analytic continuation
\begin{equation} \label{eq:analytic-cont}
\Phi \rightarrow T~, \qquad (X,Y) \rightarrow i(U,V)~, \qquad (\beta_1,\beta_2,\beta_3) \rightarrow (\alpha_1,\alpha_2,\alpha_3)~, \qquad ds^2_b \rightarrow - ds^2_a~.
\end{equation}
The scalar curvature $ R $ of the metric $ ds^2 $ is constant and given by
\begin{equation}
R=\frac{2(\alpha_1^2+\alpha_2^2+\alpha_3^2)}{\alpha_1\alpha_2\alpha_3}-4\left(\frac{1}{\alpha_1}+\frac{1}{\alpha_2}+\frac{1}{\alpha_3}\right)-\frac{2(\beta_1^2+\beta_2^2+\beta_3^2)}{\beta_1\beta_2\beta_3}+4\left(\frac{1}{\beta_1}+\frac{1}{\beta_2}+\frac{1}{\beta_3}\right).
\end{equation}
Note that this vanishes when $\al_j = \be_j$ for $j=1,2,3$.
For later convenience we also introduce the alternative deformation parameters $\tg_j$ and $\ga_j$ through the relations (assuming $\alpha_j > 0$ and $\beta_j>0$)
\begin{equation}
\begin{aligned}
\tg_1 &= \frac{\al_2}{\sqrt{\al_1\al_2\al_3}}~,
&\qquad
\tg_2 &= \frac{\al_1-\al_3}{\sqrt{\al_1\al_2\al_3}}~,
&\qquad
\tg_3 &= \frac{\al_1 - \al_2 + \al_3}{\sqrt{\al_1\al_2\al_3}}~, \\
\gamma_1 &= \frac{\be_2}{\sqrt{\be_1\be_2\be_3}}~,
&\qquad
\gamma_2 &= \frac{\be_1-\be_3}{\sqrt{\be_1\be_2\be_3}}~,
&\qquad
\gamma_3 &= \frac{\be_1 - \be_2 + \be_3}{\sqrt{\be_1\be_2\be_3}}~.
\label{eq:gammas}
\end{aligned}
\end{equation}
Our nomenclature for the different types of deformations is then defined according to \tabref{table:deformations}.
\begin{table}[t]
\centering
\begin{tabular}{c|c|c}
$ \{\alpha_1,\alpha_2,\alpha_3\} $ & $ \{\tg_1,\tg_2,\tg_3\} $ & Deformation type \\ \hline
$ \alpha_1\neq \alpha_2\neq \alpha_3 $ & $ \tg_1\neq \tg_2 \neq \tg_3 $ & elliptic \\
$ \alpha_1= \alpha_3\neq \alpha_2 $ & $ \tg_1\neq\tg_3,\, \tg_2=0 $ & trigonometric \\
$ \alpha_1=\alpha_2=\alpha_3 $ & $ \tg_1=\tg_3,\, \tg_2=0 $ & rational (undeformed)
\end{tabular}
\caption{Definition of the rational (undeformed), trigonometric and elliptic deformations.}
\label{table:deformations}
\end{table}

\subsection{Remarks on the chosen parametrisation}

When $\alpha_j=\beta_j=1$ for $j=1,2,3$ the action \eqref{eq:Lag-sigma-model} corresponds to the PCM on $\grp{SL}(2;\Real) \times \grp{SU}(2)$, and we recover the metric of $\AdS_3 \times \Sp^3$, albeit in a perhaps unfamiliar form,
\begin{equation} \label{eq:metric-undef} \begin{aligned}
ds_0^2 &= -dT^2 + dU^2 + dV^2 - 2 \sinh 2 U \, dT dV + d \Phi^2 + d Y^2 + d X^2 - 2 \sin 2X d \Phi d Y~.
\end{aligned}
\end{equation}
This metric follows from the usual embedding into $\Real^{2,2}$ and $\Real^4$ with
\begin{equation} \label{eq:embedding-1} \begin{aligned}
\mathcal U_0 &= \cosh U \cosh V \cos T - \sinh U \sinh V \sin T~, &\quad \mathcal X_1 &= \cos X \cos Y \cos \Phi + \sin X \sin Y \sin \Phi~, \\
\mathcal U_1 &= \cosh U \cosh V \sin T + \sinh U \sinh V \cos T~, &\quad \mathcal X_2 &= \cos X \cos Y \sin \Phi - \sin X \sin Y \cos \Phi~, \\
\mathcal U_2 &= \cosh U \sinh V \sin T + \sinh U \cosh V \cos T~, &\quad
\mathcal X_3 &= \cos X \sin Y \sin \Phi + \sin X \cos Y \cos \Phi~, \\
\mathcal U_3 &= \cosh U \sinh V \cos T - \sinh U \cosh V \sin T~, &\quad \mathcal X_4 &= \cos X \sin Y \cos \Phi - \sin X \cos Y \sin \Phi~,
\end{aligned}
\end{equation}
satisfying
\begin{equation} \label{eq:embedding-constraints}
\begin{aligned}
-\mathcal U_0^2-\mathcal U_1^2+\mathcal U_2^2+\mathcal U_3^2 &= -1~, &\qquad -d\mathcal U_0^2-d\mathcal U_1^2+d\mathcal U_2^2+d\mathcal U_3^2 &= ds_{a,0}^2~, \\
\mathcal X_1^2+\mathcal X_2^2+\mathcal X_3^2+\mathcal X_4^2 &= +1~, &\qquad d\mathcal X_1^2+d\mathcal X_2^2+d\mathcal X_3^2+d\mathcal X_4^2 &= ds_{s,0}^2~.
\end{aligned}
\end{equation}
A possible solution of the equations of motion and Virasoro constraints for a point-like ($\sigma$-independent) string propagating in the background \eqref{eq:metric-undef} is given by
\begin{equation} \label{eq:cs-1}
T=\Phi=\tau~, \qquad U=V=X=Y=0~.
\end{equation}
On this classical solution the embedding coordinates are
\begin{equation} \label{eq:gauge}
\begin{aligned}
\mathcal U_0 &= \cos \tau~, &\qquad \mathcal U_1 &= \sin \tau~, &\qquad \mathcal U_2 &= \mathcal U_3=0~, \\
\mathcal X_1 &= \cos \tau~, &\qquad \mathcal X_2 &=\sin \tau~, &\qquad \mathcal X_3 &=\mathcal X_4 =0~.
\end{aligned}
\end{equation}
At this point it is useful to note that instead of \eqref{eq:param} we could have chosen another parametrisation
\begin{equation}
\tilde{g} = e^{\frac{i}{2}(t + \psi) L_3} e^{\text{arcsinh} \rho \, L_1} e^{\frac{i}{2} (t-\psi) L_3} e^{\frac{1}{2}(\phi+\varphi) J_3} e^{\arcsin r \,J_1} e^{\frac{1}{2} (\phi-\varphi) J_3}~,
\end{equation}
which in the undeformed case leads to the metric of $\AdS_3 \times \Sp^3$ in global coordinates,
\begin{equation} \label{eq:metric-global}
\begin{aligned}
d\tilde{s}_{0}^2 &= - (1+\rho^2) dt^2 + \frac{d\rho^2}{1+\rho^2} + \rho^2 d\psi^2 + (1-r^2) d\phi^2 + \frac{dr^2}{1-r^2} + r^2 d\varphi^2~.
\end{aligned}
\end{equation}
In terms of
\begin{equation}
u = \rho \cos \psi~, \qquad v = \rho \sin \psi~, \qquad x = r \cos \varphi~, \qquad y = r \sin \varphi~,
\end{equation}
the analytic continuation relating the metric of $\AdS_3$ and $\Sp^3$ in \eqref{eq:metric-global} is
\begin{equation}
\phi \rightarrow t~, \qquad (x,y) \rightarrow i(u,v)~, \qquad d \tilde{s}_b^2 \rightarrow d\tilde{s}_a^2~,
\end{equation}
while a possible choice for the embedding coordinates is
\begin{equation} \label{eq:embedding-2}
\begin{aligned}
\tilde{\mathcal U}_0 &= \sqrt{1+u^2+v^2} \cos t~, &\qquad \tilde{\mathcal X}_1 &= \sqrt{1-x^2-y^2} \cos \phi~, \\
\tilde{\mathcal U}_1 &= \sqrt{1+u^2+v^2} \sin t~, &\qquad \tilde{\mathcal X}_2 &= \sqrt{1-x^2-y^2} \sin \phi~, \\
\tilde{\mathcal U}_2 &= u~, &\qquad \tilde{\mathcal X}_3 &= x~, \\
\tilde{\mathcal U}_3 &= v~, &\qquad \tilde{\mathcal X}_4 &= y~,
\end{aligned}
\end{equation}
satisfying the analogous relations to \eqref{eq:embedding-constraints}.
Moreover, on the classical solution
\begin{equation} \label{eq:cs-2}
t=\phi=\tau~, \qquad u=v=x=y=0~,
\end{equation}
the embedding coordinates are again given by \eqref{eq:gauge} (now with tilded quantities on the left-hand sides).
Equating the embedding coordinates \eqref{eq:embedding-1} and \eqref{eq:embedding-2} one finds the change of variables
\begin{equation} \label{eq:change-var}
\begin{aligned}
t &= \arctan \left( \frac{\sin T + \tanh U \tanh V \cos T}{\cos T - \tanh U \tanh V \sin T} \right)~, &\qquad \phi &= \arctan \left( \frac{\sin \Phi - \tan X \tan Y \cos \Phi}{\cos \Phi + \tan X \tan Y \sin \Phi} \right)~, \\
u &= \cosh U \sinh V \sin T + \sinh U \cosh V \cos T~, &\qquad
x &= \cos X \sin Y \sin \Phi + \sin X \cos Y \cos \Phi~, \\
v &= \cosh U \sinh V \cos T - \sinh U \cosh V \sin T~, &\qquad
y &= \cos X \sin Y \cos \Phi - \sin X \cos Y \sin \Phi~, \\
\end{aligned}
\end{equation}
which maps the classical solution \eqref{eq:cs-2} to the classical solution \eqref{eq:cs-1}.
Moreover, for small excitations $(u,v)$ and $(x,y)$, and setting $t=T=\phi=\Phi=\tau$ the relation between the transverse coordinates is a $\tau$-dependent rotation
\begin{equation}
\begin{pmatrix} u \\ v \end{pmatrix}
=
\begin{pmatrix}
\cos \tau & \sin \tau \\
-\sin \tau & \cos \tau
\end{pmatrix}
\begin{pmatrix}
U \\ V
\end{pmatrix}~, \qquad
\begin{pmatrix} x \\ y \end{pmatrix}
=
\begin{pmatrix}
\cos \tau & \sin \tau \\
-\sin \tau & \cos \tau
\end{pmatrix}
\begin{pmatrix}
X \\ Y
\end{pmatrix}~.
\end{equation}
In later sections, when computing the S-matrix describing the scattering of the transverse excitations $U,V,X,Y$, we will analyse how this $\tau$-dependent rotation of the excitations affects the scattering.
Let us note that this change of variables also applies when considering the trigonometric deformation.
Indeed, in the case
\begin{equation}
\al_1 = \al_3 = \be_1 = \be_3 = \frac{1}{1+\kappa^2}~, \qquad \al_2=\be_2=1~,
\end{equation}
implementing the change of variables \eqref{eq:change-var} in the metric \eqref{eq:ellipticmetricAdS3xS3} gives
\begin{equation} \label{eq:metric-YB} \begin{aligned}
ds^2_\kappa &= \frac{1}{1+\kappa^2} \left(-(1+\rho^2) dt^2 + \frac{d\rho^2}{1+\rho^2} + \rho^2 d \psi^2 - \kappa^2 ((1+\rho^2) dt - \rho^2 d\psi)^2 \right) \\
&\qquad +\frac{1}{1+\kappa^2} \left((1-r^2) d\phi^2 + \frac{dr^2}{1-r^2} + r^2 d \varphi^2 + \kappa^2 ((1-r^2) d\phi + r^2 d\varphi)^2 \right)~,
\end{aligned}
\end{equation}
which is simply the metric of the unilateral inhomogeneous YB deformation of $\AdS_3 \times \Sp^3$~\cite{Klimcik:2002zj,Hoare:2014oua}.

\section{Light-cone gauge-fixed theory}\label{sec:lcgft}

To quantise the model and find the S-matrix it is necessary to remove the redundancies in the definition of the action.
In particular, the sigma model action is invariant under reparametrisations of the string worldsheet.
In the context of strings on $\AdS$ spaces, this redundancy is typically removed by fixing uniform light-cone gauge~\cite{Arutyunov:2005hd}.
In this section we briefly review this light-cone gauge fixing, both in the Hamiltonian and Lagrangian formalism.
For further details the reader is invited to consult the review~\cite{Arutyunov:2009ga} as well as the paper~\cite{Arutyunov:2014jfa}.
We then write down the gauge-fixed Lagrangians for the elliptically-deformed model, expanded to quartic order in the transverse fields.

\subsection{Classical solution}

The first step in fixing uniform light-cone gauge is to identify a classical solution around which the action can be expanded.
For strings propagating in a background that is invariant under shifts of two coordinates, this classical solution is usually a point-like string with the two isometric directions identified with the worldsheet time coordinate $\tau$.
The only isometries left after the elliptic deformation are shifts in $T$ and $\Phi$, associated to the left Cartan generators.
These can then be used to fix uniform light-cone gauge.
Note that this does not correspond to the standard gauge fixing for strings in (undeformed) $\AdS_3$.
There, it is usually the global time $t$ and the angle $\varphi$ in the $S^3$, corresponding to the ``diagonal'' Cartan generators, that are used for light-cone gauge fixing.
As we will see, this alternative light-cone gauge fixing, which is forced upon us by symmetries, leads to a modified dispersion relation for the excitations.
For later convenience (in particular to work with canonically-normalised quantities), we find it convenient to assume positive deformation parameters
\begin{equation}
\be_j > 0~, \qquad \al_j >0~,
\end{equation}
and rescale the fields as
\begin{equation}
T \rightarrow \frac{T}{\sqrt{\al_2}}~, \qquad U \rightarrow \frac{U}{\sqrt{\al_3}}~, \qquad V \rightarrow \frac{V}{\sqrt{\al_1}}~, \qquad \Phi \rightarrow \frac{\Phi}{\sqrt{\be_2}}~, \qquad X \rightarrow \frac{X}{\sqrt{\be_3}}~, \qquad Y \rightarrow \frac{Y}{\sqrt{\be_1}}~.
\end{equation}
One can then check that the point-like string parametrised by
\begin{equation}
T = \Phi = \lambda \tau~, \qquad U=V=X=Y=0~, \qquad \lambda \in \Real~,
\end{equation}
solves the equations of motion and the Virasoro constraints, including for the elliptically-deformed model.

\subsection{Uniform light-cone gauge}

We introduce the target-space light-cone coordinates
\begin{equation}
X^+ = (1-a) T + a \Phi~, \qquad X^- = - T + \Phi~, \qquad T = X^+ - a X^-~, \qquad \Phi = X^+ + (1-a) X^-~,
\end{equation}
with free parameter $a \in [0,1]$.
In the Hamiltonian formalism, to fix uniform light-cone gauge we introduce the conjugate momenta
\begin{equation}
P_+ = P_T + P_\Phi~, \qquad P_- = -a P_T + (1-a) P_\Phi~, \qquad P_T = (1-a) P_+ - P_-~, \qquad P_\Phi = a P_+ + P_-~,
\end{equation}
obtained from the sigma model action \eqref{eq:Lag-sigma-model} in the usual way
\begin{equation}
P_\mu = \frac{\delta \,\Act}{\delta \,\partial_\tau \field^\mu} = - \gamma^{\tau \beta} G_{\mu \nu} \partial_\beta \field^\nu~,
\end{equation}
where we have set the B-field to vanish, and set
\begin{equation}
X^+ = \tau~, \qquad P_- = 1~.
\end{equation}
The light-cone gauge-fixed Hamiltonian is then given by
\begin{equation}
\Ham^{\text{g.f.}} = - P_+(\field^j, \partial_\sigma \field^j, P_j)~,
\end{equation}
obtained by solving the Virasoro constraints
\begin{align}
P_\mu \partial_\sigma \field^\mu = 0~, \qquad
G^{\mu \nu} P_\mu P_\nu + G_{\mu \nu} \partial_\sigma \field^\mu \partial_\sigma \field^\nu = 0~.
\end{align}
The index $j$ runs over all target-space coordinates except those involved in the light-cone gauge fixing.
In our case we have $\field^j = (U, V, X, Y)^j$.

There is an equivalent way to obtain the light-cone gauge-fixed theory, working exclusively in the Lagrangian formalism \cite{Kruczenski:2004cn,Zarembo:2009au}.
The procedure is to T-dualise the sigma model in $X^-$, and set
\begin{equation}
X^+ = \tau~, \qquad \mathring{X}^- = \sigma~,
\end{equation}
where $\mathring{X}^-$ is the T-dual coordinate.
Recalling that we specialise to the case of vanishing B-field, the light-cone gauge-fixed Lagrangian is then given by
\begin{equation}
\Lag^{\text{g.f.}} = 2 \sqrt{-\det \left[ \mathring{G}_{\mu \nu} \partial_\alpha \field^\mu \partial_\beta \field^\nu \right]}- \epsilon^{\alpha \beta} \mathring{B}_{\mu \nu} \partial_\alpha \Psi^\mu \partial_\beta \Psi^\nu ~,
\end{equation}
where here $\field^\mu = (X^+,\mathring{X}^-,U,V,X,Y)^\mu$, and
\begin{equation}
\begin{aligned}
\mathring{G}_{--} &= \frac{1}{G_{--}}~, &\qquad \mathring{G}_{- \bar{\mu}} &= \mathring{G}_{ \bar{\mu} -} = 0~, &\qquad \mathring{G}_{\bar{\mu}\bar{\nu}} &= G_{\bar{\mu} \bar{\nu}} - \frac{G_{-\bar{\mu}} G_{-\bar{\nu}}}{G_{--}}~,\\
& & \mathring{B}_{-\bar{\mu}} &= - \mathring{B}_{\bar{\mu}-} = - \frac{G_{- \bar{\mu}}}{G_{--}}~, &\qquad \mathring{B}_{\bar{\mu}\bar{\nu}} &= 0~,
\end{aligned}
\end{equation}
and the index $\bar{\mu}$ runs over all coordinates except the one involved in the T-duality.

The light-cone gauge-fixed Lagrangian and Hamiltonian are related through the Legendre transform
\begin{equation}
\Ham^{\text{g.f.}} = P_j \partial_\tau \field^j - \Lag^{\text{g.f.}}~, \qquad P_j = \frac{\partial \Lag^{\text{g.f.}}}{\partial(\partial_\tau \field^j)}~.
\end{equation}
To obtain the S-matrix we need to expand the gauge-fixed Lagrangian or Hamiltonian up to quartic order in the transverse fields.
It turns out that for the action \eqref{eq:Lag-sigma-model} there are no odd terms in the expansion
\begin{equation}
\Lag^{\text{g.f.}} = \Lag_2^{\text{g.f.}} + \Lag_4^{\text{g.f.}} + \dots~, \qquad \Ham^{\text{g.f.}} = \Ham_2^{\text{g.f.}} + \Ham_4^{\text{g.f.}} + \dots~.
\end{equation}
Henceforth, for convenience, we remove the $^{\text{g.f.}}$ superscript.
In the next subsection the Lagrangians and Hamiltonians will always be those of the light-cone gauge-fixed theory.

\subsection{Light-cone gauge-fixed Lagrangians}

Fixing light-cone gauge around the classical solution $ X=Y=U=V=0 $ as described in the previous subsection gives the following quadratic Lagrangian
\begin{equation}
\begin{aligned}
\mathcal{L}_2&=
\frac{1}{2}\left(\dot{U}^2-\acute{U}^2+\dot{V}^2-\acute{V}^2\right)+\frac{2(\al_1-\al_2)}{\al_2\al_3}U^2-\frac{2(\al_2-\al_3)}{\al_1\al_2}V^2-2\frac{\sqrt{\al_1}}{\sqrt{\al_2\al_3}}U\dot{V}-\frac{2(\al_2-\al_3)}{\sqrt{\al_1\al_2\al_3}}\dot{U}V\\
&+\frac{1}{2}\left(\dot{X}^2-\acute{X}^2+\dot{Y}^2-\acute{Y}^2\right)+\frac{2(\be_1-\be_2)}{\be_2\be_3}X^2-\frac{2(\be_2-\be_3)}{\be_1\be_2}Y^2-2\frac{\sqrt{\be_1}}{\sqrt{\be_2\be_3}}X\dot{Y}-\frac{2(\be_2-\be_3)}{\sqrt{\be_1\be_2\be_3}}\dot{X}Y~,\\
\end{aligned}
\label{eq:L2ellipticAdS3}
\end{equation}
describing four real fields whose equations of motion are given by
\begin{equation}
\begin{aligned}
& \ddot{U}-U^{\prime\prime}-4\frac{(\al_1-\al_2)}{\al_2\al_3}U+\frac{2(\al_1-\al_2+\al_3)}{\sqrt{\al_1\al_2\al_3}}\dot{V}=0~,\\
& \ddot{V}-V^{\prime\prime}+4\frac{(\al_2-\al_3)}{\al_1\al_2}V-\frac{2(\al_1-\al_2+\al_3)}{\sqrt{\al_1\al_2\al_3}}\dot{U}=0~,\\
& \ddot{X}-X^{\prime\prime}-4\frac{(\be_1-\be_2)}{\be_2\be_3}X+\frac{2(\be_1-\be_2+\be_3)}{\sqrt{\be_1\be_2\be_3}}\dot{Y}=0~,\\
& \ddot{Y}-Y^{\prime\prime}+4\frac{(\be_2-\be_3)}{\be_1\be_2}Y-\frac{2(\be_1-\be_2+\be_3)}{\sqrt{\be_1\be_2\be_3}}\dot{X}=0~.
\label{eq:EOMellipticAdS3}
\end{aligned}
\end{equation}
At this point it is convenient to introduce complex variables
\begin{equation}
W = \frac{1}{\sqrt{2}}(U + i V)~, \qquad \overline{W} = \frac{1}{\sqrt{2}}(U - i V)~, \qquad Z = \frac{1}{\sqrt{2}}(X + i Y)~, \qquad \overline{Z} = \frac{1}{\sqrt{2}}(X - i Y)~.
\label{eq:changeofvar}
\end{equation}
The quadratic light-cone gauge-fixed Lagrangian $ \Lag_2 $ can then be rewritten, up to total derivatives, as
\begin{equation}
\begin{aligned}
\mathcal{L}_2=& |\dot{W}|^2-|\acute{W}|^2-i\tg_3\left(W\dot{\overline{W}}-\dot{W}\overline{W}\right)-(\tg_1^2-\tg_2^2-\tg_3^3)|W|^2+\tg_2\tg_3(W^2+\overline{W}^2)\\
&+|\dot{Z}|^2-|\acute{Z}|^2-i\gamma_3\left(Z\dot{\overline{Z}}-\dot{Z}\overline{Z}\right) -(\gamma_1^2-\gamma_2^2-\gamma_3^3)|Z|^2+\gamma_2\gamma_3(Z^2+\overline{Z}^2)~,
\end{aligned}
\label{eq:Lag2elliptic}
\end{equation}
where the alternative deformation parameters $ \{\ga_j,\tg_j\} $, $ j=1,2,3 $ were introduced in eq.~\eqref{eq:gammas}, while the equations of motion become
\begin{equation}
\begin{aligned}
\mathcal E_W &\coloneqq \ddot{W}-W^{\prime\prime}-2i\tg_3\dot{W}+(\tg_1^2-\tg_2^2-\tg_3^2)W-2\tg_2\tg_3\overline{W}=0~,\\
\mathcal E_{\overline{W}} &\coloneqq \ddot{\overline{W}}-\overline{W}^{\prime\prime}+2i\tg_3\dot{\overline{W}}+(\tg_1^2-\tg_2^2-\tg_3^2)\overline{W}-2\tg_2\tg_3W=0~,\\
\mathcal E_Z &\coloneqq \ddot{Z}-Z^{\prime\prime}-2i\gamma_3\dot{Z}+(\gamma_1^2-\gamma_2^2-\gamma_3^2)Z-2\gamma_2\gamma_3\overline{Z}=0~,\\
\mathcal E_{\overline{Z}} &\coloneqq \ddot{\overline{Z}}-\overline{Z}^{\prime\prime}+2i\gamma_3\dot{\overline{Z}}+(\gamma_1^2-\gamma_2^2-\gamma_3^2)\overline{Z}-2\gamma_2\gamma_3Z=0~.
\end{aligned}
\label{eq:EOM-complex}
\end{equation}
The first two equations $\mathcal E_W$ and $\mathcal E_{\overline{W}}$ as well as the last two equations $\mathcal E_Z$ and $\mathcal E_{\overline{Z}}$ are coupled differential equations.
They become decoupled in the special case $\tg_2 \tg_3 = \ga_2 \ga_3 =0$.

Contrary to $ \Lag_2 $, the quartic interaction Lagrangian $ \mathcal{L}_4 $ not only has terms depending solely on the $ \AdS_3 $ coordinates $ \{W,\overline{W}\}$ and solely on the $ \Sp^3 $ coordinates $ \{Z,\overline{Z}\} $, but it also has mixed terms depending on all four fields $ \{W,\overline{W},Z,\overline{Z}\} $.
In order to write $ \mathcal{L}_4 $ in a simple form let us start by defining the $ T\overline{T} $-operator
\begin{equation}
O_{T \bar{T}} = T_{\tau \tau} T_{\sigma \sigma} - T_{\tau\sigma} T_{\sigma \tau}~,
\label{eq:TTbar}
\end{equation}
where $ T_{\mu\nu} $ denotes the energy-momentum tensor.
The energy-momentum tensor gathers the conserved currents associated to translation invariance on the worldsheet of the string, and, to quadratic order, can be computed using the standard formula
\begin{equation}
T^{\mu}{}_\nu = \frac{\partial \mathcal L_2}{\partial_\mu \field^j} \partial_\nu \field^j - \delta^{\mu}_\nu \mathcal L_2~,
\end{equation}
where the sum is over all the fields $\field^j=(W,\overline{W},Z,\overline{Z})^j$.
Using the quadratic Lagrangian in eq.~\eqref{eq:Lag2elliptic} we then find the explicit expressions
\begin{equation}
\begin{aligned}
& T_{\tau\tau}=-|\dot{W}|^2-|\acute{W}|^2-\left(\tg_1^2-\tg_2^2-\tg_3^2\right)|W|^2+\tg_2\tg_3(W^2+\overline{W}^2)\\[0.5ex]
& \hspace{0.95cm} -|\dot{Z}|^2-|\acute{Z}|^2-\left(\ga_1^2-\ga_2^2-\ga_3^2\right)|Z|^2+\ga_2\ga_3(Z^2+\overline{Z}^2)~,\\[0.7ex]
& T_{\sigma\sigma}=-|\dot{W}|^2-|\acute{W}|^2+\left(\tg_1^2-\tg_2^2-\tg_3^2\right)|W|^2-\tg_2\tg_3(W^2+\overline{W}^2)+i\ga_3(W\dot{\bar{W}}-\dot{W}\bar{W})\\[0.5ex]
& \hspace{0.95cm} -|\dot{Z}|^2-|\acute{Z}|^2+\left(\ga_1^2-\ga_2^2-\ga_3^2\right)|Z|^2-\ga_2\ga_3(Z^2+\overline{Z}^2)+i\ga_3(Z\dot{\overline{Z}}-\dot{Z}\overline{Z})~,\\[0.7ex]
&T_{\sigma \tau}=-\dot{W}\acute{\overline{W}}-\acute{W}\dot{\overline{W}}-\dot{Z}\acute{\overline{Z}}-\acute{Z}\dot{\overline{Z}}~,\\[0.7ex]
&T_{\tau \sigma}=-\dot{W}\acute{\overline{W}}-\acute{W}\dot{\overline{W}}-\dot{Z}\acute{\overline{Z}}-\acute{Z}\dot{\overline{Z}}+i\tg_3(W\acute{\overline{W}}-\acute{W}\overline{W})+i\ga_3(Z\acute{\overline{Z}}-\acute{Z}\overline{Z}) ~.
\end{aligned}
\label{eq:Tmunu}
\end{equation}
With the above definitions we can write the quartic gauge-fixed Lagrangian as
\begin{equation}
\begin{aligned}
\Lag_4 & =	\check{\Lag}_4(W,\overline{W})+\hat{\Lag}_4(Z,\overline{Z})+\tilde{\Lag}_4(W,\overline{W},Z,\overline{Z})-\left(a-\frac{1}{2}\right)O_{T\bar{T}} \\
&\quad +\frac{i(a-1)}{2}(\tg_1+\tg_2)(W^2-\overline{W}^2)\left(\dot{Z}\mathcal{E}_{\overline{Z}}+\dot{\overline{Z}}\mathcal{E}_{Z}\right)+\frac{ia}{2}(\ga_1+\ga_2)(Z^2-\overline{Z}^2)\left(\dot{W}\mathcal{E}_{\overline{W}}+\dot{\overline{W}}\mathcal{E}_W\right)\\[0.5ex]
&\quad+\frac{i(a-1)}{2}(\tg_1+\tg_2)(W^2-\overline{W}^2)\left(\dot{W}\mathcal{E}_{\bar{W}}+\dot{\bar{W}}\mathcal{E}_W\right)+\frac{ia}{2}(\ga_1+\ga_2)(Z^2-\overline{Z}^2)\left(\dot{Z}\mathcal{E}_{\overline{Z}}+\dot{\overline{Z}}\mathcal{E}_Z\right)~.\\[0.5ex]
\end{aligned}
\label{eq:L4ellipticTTbarJTbar}
\end{equation}
The remaining terms containing only $ \AdS_3 $ fields are given, up to total derivatives, by
\begin{equation}
\begin{aligned}
\check{\Lag}_4 &= -2 (\tg_1^2-\tg_2^2-\tg_3^2) |W|^2 |\dot{W}|^2 + \frac{i}{2}\xi_1 |W|^2 \left( W \dot{\overline{W}} - \dot{W}\overline{W} \right) + \xi_2 |W|^4\\
&\quad -\frac{1}{2} \tg_3 \overline{W} \left( (\tg_1+\tg_3)\overline{W} - i \dot{\overline{W}} \right) \left( \dot{W}^2 - \acute{W}^2 \right)
-\frac{1}{2} \tg_3 W \left( (\tg_1+\tg_3)W + i \dot{W} \right) \left( \dot{\overline{W}}^2 - \acute{\overline{W}}^2 \right) \\
&\quad - \tg_2 \tg_3 |W|^2 \left( \dot{W}^2 - \acute{W}^2 + \dot{\overline{W}}^2 - \acute{\overline{W}}^2 \right) + 2 \tg_2 \tg_3 |\dot{W}|^2 \left( W^2 + \overline{W}^2 \right) - \frac{1}{3} \tg_2 \xi_3 \left( W^3 \overline{W} + W \overline{W}^3 \right) \\
&\quad -2 i \tg_2 \xi_6 |W|^2 \left( W \dot{W} - \overline{W} \dot{\overline{W}}\right) -\frac{1}{6} \tg_2^2 \xi_4 (W^4 + \overline{W}^4)+ \frac{1}{12} \xi_0 \left( W^3 \mathcal E_W + \overline{W}^3 \mathcal E_{\overline{W}} \right) \\
&\quad +\frac{1}{4} \xi_5 \left( \mathcal E_W \overline{W} \left(W^2 - \overline{W}^2 - |W|^2 \right) + \mathcal E_{\overline{W}} W \left(\overline{W}^2 - W^2 - |W|^2 \right) \right)~,
\end{aligned}
\label{eq:Lag4AdS3part}
\end{equation}
where the explicit form of the constants $\xi_j $, $j=0,1,...,5 $ in eq.~\eqref{eq:Lag4AdS3part} is
\begin{equation}
\begin{aligned}
\xi_0 &= \tg_1^2+2\tg_2^2+3\tg_1\tg_2+\tg_1\tg_3~, \\
\xi_1 &= 2 \tg_1^3 + 3 \tg_1^2 \tg_3 - 7 \tg_2^2 \tg_3 - 3 \tg_3^3 - 2 \tg_1(\tg_2^2+\tg_3^2) \\
\xi_2 &= \tg_2^4 - \tg_1^3 \tg_3 + 4 \tg_2^2 \tg_3^2 + \tg_3^4 -\tg_1^2 (\tg_2^2+\tg_3^2)+\tg_1\tg_3 (3 \tg_2^2 + \tg_3^2)~, \\
\xi_3 &= 2 \tg_1^3 + 3 \tg_1^2 \tg_3 - 7 \tg_2^2 \tg_3 - 2 \tg_1 \tg_2^2 - 5 \tg_1 \tg_3^2 - 6 \tg_3^3~, \\
\xi_4 &=\tg_1^2 - \tg_2^2 - 2 \tg_1 \tg_3 - 7 \tg_3^2~, \\
\xi_5 &=\tg_1 (\tg_1 + \tg_2 + \tg_3)~,\\
\xi_6 &=\tg_1^2 -\tg_2^2 - 2 \tg_1 \tg_3 - 4 \tg_3^2~.
\end{aligned}
\end{equation}
The terms containing only $ \Sp^3 $ fields can be obtained from the $ \AdS_3 $ ones in the following way,
\begin{equation}
\hat{\Lag}_4=-\check{\Lag}_4\Big|_{W\rightarrow Z,\tg_i\rightarrow \ga_i}.\label{eq:L4W}
\end{equation}
Finally, the mixed terms in eq.~\eqref{eq:L4ellipticTTbarJTbar} are given by
\begin{equation}
\begin{aligned}
\tilde{\Lag}_4&=\left((\ga_1^2-\ga_2^2-\ga_3^2)|Z|^2-\ga_2\ga_3(Z^2+\overline{Z}^2)\right)\left(|\dot{W}|^2+|\acute{W}|^2\right)\\[0.5ex]
&\quad-\left((\tg_1^2-\tg_2^2-\tg_3^2)|W|^2-\tg_2\tg_3(W^2+\overline{W}^2)\right)\left(|\dot{Z}|^2+|\acute{Z}|^2\right)\\[0.5ex]
&\quad+\frac{i}{2}\ga_3\left((\tg_1^2-\tg_2^2-\tg_3^2)|W|^2-\tg_2\tg_3(W^2+\overline{W}^2)\right)(Z\dot{\overline{Z}}-\dot{Z}\overline{Z})\\[0.5ex]
&\quad-\frac{i}{2}\tg_3\left((\ga_1^2-\ga_2^2-\ga_3^2)|Z|^2-\ga_2\ga_3(Z^2+\overline{Z}^2)\right)(W\dot{\overline{W}}-\dot{W}\overline{W})\\[0.5ex]
&\quad+\frac{i}{2}\ga_3\left(Z\dot{\overline{Z}}-\dot{Z}\overline{Z}\right)\left(\dot{W}\dot{\overline{W}}+\acute{W}\acute{\overline{W}}\right)-\frac{i}{2}\ga_3\left(Z\acute{\overline{Z}}-\acute{Z}\overline{Z}\right)\left(\dot{W}\acute{\overline{W}}+\acute{W}\dot{\overline{W}}\right)\\[0.5ex]
&\quad-\frac{i}{2}\tg_3\left(W\dot{\overline{W}}-\dot{W}\overline{W}\right)\left(\dot{Z}\dot{\overline{Z}}+\acute{Z}\acute{\overline{Z}}\right)-\frac{i}{2}\tg_3\left(W\acute{\overline{W}}-\acute{W}\overline{W}\right)\left(\dot{Z}\acute{\overline{Z}}+\acute{Z}\dot{\overline{Z}}\right)~.
\end{aligned}
\end{equation}
When $\ga_2 = \tg_2=0$ (corresponding to $\alpha_1 = \alpha_3$ and $\beta_1 = \beta_3$ in the original set of deformation parameters, hence describing both the undeformed and trigonometric deformed cases) both the quadratic and quartic Lagrangians in \eqref{eq:Lag2elliptic} and \eqref{eq:L4ellipticTTbarJTbar} are invariant under the two $\alg{u}(1)$ transformations
\begin{equation} \label{eq:u1sym}
W \rightarrow e^{i \epsilon} W~, \qquad Z \rightarrow e^{i \epsilon'} Z~.
\end{equation}
As we will see later, these additional symmetries can be used to perform a time-dependent rotation of the transverse fields and bring the quadratic and quartic gauge-fixed Lagrangians to their usual undeformed and YB-deformed form.

\section{Light-cone gauge S-matrix}\label{sec:tlsmat}

With the knowledge of the gauge-fixed Lagrangian we can compute the tree-level S-matrix.
The first step is to make a plane-wave ansatz for the fields in order to define asymptotic incoming and outgoing scattering states that solve the equations of motion of the quadratic Lagrangian $\mathcal L_2$.
The second step is to use $\Lag_4$ to deduce the interactions between these states.
As we will see, the tree-level S-matrix is diagonal and immediately satisfies the classical YB equation, which is in agreement with the fact that the elliptically-deformed model is integrable.

\subsection{Oscillators}

To solve the equations of motion \eqref{eq:EOM-complex} it is useful to go to momentum space and make the following plane-wave ansatz for the fields
\begin{align} \label{eq:o-W}
W = \frac{1}{\sqrt{2}} \int dp \left( A_+ e^{-i \tom_{+} \tau + i p \sigma} a_+ + A_- e^{-i \tom_- \tau + i p \sigma} a_- + B_+ e^{+i \tom_+ \tau - i p \sigma} a_+^\dagger + B_- e^{+i \tom_- \tau - i p \sigma} a_-^\dagger\right)~, \\
\overline{W} = \frac{1}{\sqrt{2}} \int dp \left( \overline{B}_+ e^{-i \tom_+ \tau + i p \sigma} a_+ + \overline{B}_- e^{-i \tom_- \tau + i p \sigma} a_- + \overline{A}_+ e^{+i \tom_+ \tau - i p \sigma} a_+^\dagger + \overline{A}_- e^{+i \tom_- \tau - i p \sigma} a_-^\dagger\right)~, \\
Z = \frac{1}{\sqrt{2}} \int dp \left( C_+ e^{-i \om_+ \tau + i p \sigma} b_+ + C_- e^{-i \om_- \tau + i p \sigma} b_- + D_+ e^{+i \om_+ \tau - i p \sigma} b_+^\dagger + D_- e^{+i \om_- \tau - i p \sigma} b_-^\dagger\right)~, \\
\overline{Z} = \frac{1}{\sqrt{2}} \int dp \left( \overline{D}_+ e^{-i \om_+ \tau + i p \sigma} b_+ + \overline{D}_- e^{-i \om_- \tau + i p \sigma} b_- + \overline{C}_+ e^{+i \om_+ \tau - i p \sigma} b_+^\dagger + \overline{C}_- e^{+i \om_- \tau - i p \sigma} b_-^\dagger\right)~.
\label{eq:o-Zb}
\end{align}
One can check that $\overline{W}$ and $\overline{Z}$ are complex conjugates of $W$ and $Z$ respectively.
To quantise the fields we introduce four sets of annihilation and creation operators.
For excitations in $\AdS_3$ we have the pairs $(a_+,a_+^\dagger)$ and $(a_-, a_-^\dagger)$, while for excitations in $\Sp^3$ we have $(b_+,b_+^\dagger)$ and $(b_-, b_-^\dagger)$, see also \tabref{table:oscillators} for the states created by these operators.
\begin{table}[t] \centering
\begin{tabular}{|c|c|c|c|c|c|} \hline
Oscillator & Particle & $H$ & $J^w$ & $J^z$ & Type \\ \hline
$a_+^\dagger$ & $\ket{a_+^\dagger}$ & $\tom_+$ & $+1$ & $0$ & $\AdS_3$\\ \hline
$a_-^\dagger$ & $\ket{a_-^\dagger}$ & $\tom_-$ & $-1$ & $0$ & $\AdS_3$\\ \hline
$b_+^\dagger$ & $\ket{b_+^\dagger}$ & $\om_+$ & $0$ & $+1$ & $\Sp^3$\\ \hline
$b_-^\dagger$ & $\ket{b_+^\dagger}$ & $\om_-$ & $0$ & $-1$ & $\Sp^3$ \\ \hline
\end{tabular}
\caption{Conventions for the oscillators.
The charges under the two $\alg{u}(1)$ generators refer to the rational and trigonometric limits.}
\label{table:oscillators}
\end{table}
The coefficients $(A_\pm, B_\pm, C_\pm, D_\pm)$ are fixed by requiring that the equations of motion are satisfied and that the annihilation and creation operators satisfy the canonical commutation relations.
For the fields to satisfy the equations of motion we find the constraints
\begin{equation} \label{eq:coeffs}
M_\pm^a \begin{pmatrix} A_\pm \\ \overline{B}_\pm \end{pmatrix} = 0 ~, \qquad M_\pm^b \begin{pmatrix} C_\pm \\ \overline{D}_\pm \end{pmatrix} = 0 ~,
\end{equation}
with matrices $M_\pm^a = M(\tom_\pm,\tg_j)$ and $M_\pm^b = M(\om_\pm,\ga_j)$ where
\begin{equation}
M(\omega,\gamma_j)= \begin{pmatrix} -\omega^2 + p^2 +(\gamma_1^2-\gamma_2^2-\gamma_3^2) - 2 \omega \gamma_3 & - 2 \gamma_2 \gamma_3 \\ -2 \gamma_2 \gamma_3 & -\omega^2 + p^2 +(\gamma_1^2-\gamma_2^2-\gamma_3^2) + 2 \omega \gamma_3 \end{pmatrix}~.
\end{equation}
For this system of equations to have a solution we must have $\det M_\pm^a = \det M_\pm^b = 0$, which leads to the dispersion relations (we are free to exchange the definition of $\omega_+$ and $\omega_-$)
\begin{equation}
\sqrt{\tom_\pm(p)^2 + \tg_2^2} = \sqrt{p^2+\tg_1^2} \pm \tg_3~, \qquad \sqrt{\om_\pm(p)^2 + \ga_2^2} = \sqrt{p^2+\ga_1^2} \pm \ga_3~.
\label{eq:dispersionrelationselliptic}
\end{equation}
For these expressions to be well-defined we have assumed we are in the regime where
\begin{equation} \begin{aligned}
\qquad &0 < (\al_1,\al_3) \leq \al_2~, &\qquad &0 < (\be_1,\be_3) \leq \be_2~, \\ &\ga_1^2 \geq \ga_3^2~, \quad (\ga_1 \pm \ga_3)^2 \geq \ga_2^2~, &\qquad &\tilde{\ga}_1^2 \geq \tilde{\ga}_3^2~, \qquad (\tg_1\pm\tg_3)^2 \geq \tg_2^2~.
\end{aligned}
\end{equation}
Using the dispersion relations \eqref{eq:dispersionrelationselliptic}, the normalisation coefficients in eqs.~\eqref{eq:o-W,-,eq:o-Zb} are obtained by solving the equations in \eqref{eq:coeffs} and imposing that the creation and annihilation operators satisfy the canonical commutation relations.
Up to phases this fixes
\begin{align}
A_\pm &= \overline{A}_\pm = \frac{\sqrt{\sqrt{\omega^a_\pm(p)^2 + \tilde{\ga}_2^2} - \tilde{\ga}_2} \mp \sqrt{\sqrt{\omega^a_\pm(p)^2 + \tilde{\ga}_2^2} + \tilde{\ga}_2}}{2 \sqrt{\omega^a_\pm(p)} \sqrt{\bar{\omega}^a(p)}}~,
\\
B_\pm &= \overline{B}_\pm = \frac{\pm \sqrt{\sqrt{\omega^a_\pm(p)^2 + \tilde{\ga}_2^2} - \tilde{\ga}_2} + \sqrt{\sqrt{\omega^a_\pm(p)^2 + \tilde{\ga}_2^2} + \tilde{\ga}_2}}{2 \sqrt{\omega_\pm^a(p)} \sqrt{\bar{\omega}^a(p)}}~,\\
C_\pm &= \overline{C}_\pm = \frac{\sqrt{\sqrt{\omega^b_\pm(p)^2 + \ga_2^2} - \ga_2} \mp \sqrt{\sqrt{\omega^b_\pm(p)^2 + \ga_2^2} + \ga_2}}{2 \sqrt{\omega^b_\pm(p)} \sqrt{\bar{\omega}^b(p)}}~,
\\
D_\pm &= \overline{D}_\pm = \frac{\pm \sqrt{\sqrt{\omega^b_\pm(p)^2 + \ga_2^2} - \ga_2} + \sqrt{\sqrt{\omega^b_\pm(p)^2 + \ga_2^2} +\ga_2}}{2 \sqrt{\omega^b_\pm(p)} \sqrt{\bar{\omega}^b(p)}}~,
\end{align}
where
\begin{equation}
\bar{\omega}^a(p) = \sqrt{p^2 + \tilde{\ga}_1^2}~, \qquad \bar{\omega}^b(p) = \sqrt{p^2 + \ga_1^2}~.
\end{equation}
Note that the coefficients have a non-trivial dependence on the momentum.
The oscillator representation of the conjugate momenta follows from the definitions
\begin{equation}
P_W = \dot{\overline{W}} + i \tilde{\ga}_3 \overline{W}~, \qquad P_{\overline{W}} = \dot{W} - i \tilde{\ga}_3 W~, \qquad P_Z = \dot{\overline{Z}} + i \ga_3 \overline{Z}~, \qquad P_{\overline{Z}} = \dot{Z}- i \ga_3 Z~.
\end{equation}
We can then check that the quadratic Hamiltonian takes the canonical form
\begin{equation}
\begin{aligned}
H_2
&=\int dp \left(\tom_+(p) a^\dagger_+(p) a_+(p) + \tom_-(p) a^\dagger_-(p) a_-(p) + \om_+(p) b^\dagger_+(p) b_+(p) + \om_-(p) b^\dagger_-(p) b_-(p) \right)~,
\end{aligned}
\end{equation}
and proceed to write the quartic Hamiltonian in terms of oscillators.
It turns out that, upon imposing the momentum and energy conservation, the only surviving terms are of the form
\begin{equation} \begin{aligned}
\int d\tau H_4 &= \sum_{I,J,K,L} \int dp_1 dp_2 \mathcal T_{I J}^{K L} \hat{a}_\ind{L}^\dagger (p_2) \hat{a}_K^\dagger (p_1) \hat{a}_J (p_2) \hat{a}_I (p_1)~,
\end{aligned}
\end{equation}
where $I,J,K,L$ label the type of particle (with $\hat{a}_I \in \{a_\pm,b_\pm\}$).
$\mathcal T_{IJ}^{KL}$ is then the tree-level S-matrix, obtained from the exact S-matrix through the expansion
\begin{equation}
\mathcal S = 1 + i \mathcal T + \cdots ~,
\end{equation}
where, as in the rest of the paper, we omit the overall coupling or string tension.

\subsection{Tree-level S-matrix}

For the $\AdS_3 \times \Sp^3$ superstring and its YB deformation, the bosonic tree-level worldsheet S-matrix in the standard light-cone gauge is diagonal \cite{Sundin:2013ypa,Seibold:2021lju,Bocconcello:2020qkt}. 
After elliptically deforming and using our non-standard choice of gauge, the scattering remains diagonal with the only non-trivial elements given by 
\begin{align} \label{eq:Tmat-first}
\mathcal T \ket{a_{\mu_1}^\dagger(p_1) a_{\mu_2}^\dagger(p_2)} &= (- 2 \mathcal A_{\mu_1 \mu_2}^{aa} - \mathcal B_{\mu_1 \mu_2}^{aa}- \mathcal D_{\mu_1 \mu_2}^{aa}) \ket{a_{\mu_1}^\dagger(p_1) a_{\mu_2}^\dagger(p_2)}~, \\
\mathcal T \ket{b_{\mu_1}^\dagger(p_1) b_{\mu_2}^\dagger(p_2)} &= (+ 2 \mathcal A_{\mu_1 \mu_2}^{bb} + \mathcal B_{\mu_1 \mu_2}^{bb}- \mathcal D_{\mu_1 \mu_2}^{bb}) \ket{b_{\mu_1}^\dagger(p_1) b_{\mu_2}^\dagger(p_2)}~, \\
\mathcal T \ket{a_{\mu_1}^\dagger(p_1) b_{\mu_2}^\dagger(p_2)} &= (+2 \mathcal G_{\mu_1 \mu_2}^{ab} -\mathcal D_{\mu_1 \mu_2}^{ab} ) \ket{a_{\mu_1}^\dagger(p_1) b_{\mu_2}^\dagger(p_2)}~, \\
\mathcal T \ket{b_{\mu_1}^\dagger(p_1) a_{\mu_2}^\dagger(p_2)} &= (-2 \mathcal G_{\mu_1 \mu_2}^{ba} -\mathcal D_{\mu_1 \mu_2}^{ba}) \ket{b_{\mu_1}^\dagger(p_1) a_{\mu_2}^\dagger(p_2)}~,
\label{eq:Tmat-last}
\end{align}
with coefficients
\begin{align}
\mathcal A_{\mu_1 \mu_2}^{c_1 c_2} &= \frac{1}{4} \frac{p_1^2 \omega^{c_2}_{\mu_2}(p_2)^2 + p_2^2 \omega^{c_1}_{\mu_1}(p_1)^2-2 p_1^2 p_2^2}{p_1 \omega^{c_2}_{\mu_2}(p_2) - p_2 \omega^{c_1}_{\mu_1}(p_1)}~, \\
\mathcal G_{\mu_1 \mu_2}^{c_1 c_2} &= \frac{1}{4} \left(p_1 \omega_{\mu_2}^{c_2}(p_2) + p_2 \omega_{\mu_1}^{c_1}(p_1)\right)~, \\
\mathcal D_{\mu_1 \mu_2}^{c_1 c_2} &= \left(a-\frac{1}{2} \right)(p_1 \omega_{\mu_2}^{c_2}(p_2) - p_2 \omega_{\mu_1}^{c_1}(p_1))~,
\end{align}
and
\begin{align}
\mathcal B_{\mu_1 \mu_2}^{aa} &= \mu_1 \mu_2 \left( (\tg_1 -\tg_3)^2-\tg_2^2 \right) \frac{(\sqrt{p_1^2+\tg_1^2}- \mu_1 \tg_1)(\sqrt{p_2^2+\tg_1^2}- \mu_2 \tg_1)}{p_1 \omega_{\mu_2}^{a}(p_2) - p_2 \omega^{a}_{\mu_1}(p_1)} ~,\\
\mathcal B_{\mu_1 \mu_2}^{bb} &= \mu_1 \mu_2 \left( (\ga_1-\ga_3)^2 - \ga_2^2 \right) \frac{(\sqrt{p_1^2+\ga_1^2}- \mu_1 \ga_1)(\sqrt{p_2^2+\ga_1^2}- \mu_2 \ga_1)}{p_1 \omega_{\mu_2}^{b}(p_2) - p_2 \omega_{\mu_1}^{b}(p_1)} ~.
\label{eq:Belements}
\end{align}
In the tree-level S-matrix, only the combination $\mathcal A + \frac{1}{2} \mathcal B$ appears.
In $\mathcal A$, $\mathcal G$ and $\mathcal D$ the deformation parameters and the charges of the excitations only appear implicitly through the dispersion relation, while $\mathcal B$ depends explicitly on $\ga_j, \tg_j$ and $\mu_1, \mu_2$.
Because the scattering is diagonal, the classical YB equation
\begin{equation} \label{eq:cYBE}
\com{\mathcal T_{23}}{\mathcal T_{13}} + \com{\mathcal T_{23}}{\mathcal T_{12}} + \com{\mathcal T_{13}}{\mathcal T_{12}} =0 ~,
\end{equation}
is automatically satisfied.
In eq.~\eqref{eq:cYBE} the indices denote the embedding of the scattering matrix in the tensor product space
\begin{equation*}
\mathcal T_{12} = \mathcal T \otimes 1~, \qquad \mathcal T_{23} = 1 \otimes \mathcal T~, \qquad \mathcal T_{13} = P_{23} \mathcal T_{12} P_{23} = P_{12} \mathcal T_{23} P_{23}~,
\end{equation*}
where $P_{jk}$ is the permutation operator.

A second consequence of the diagonal scattering is that the tree-level S-matrix is invariant under a $\mathfrak{u}(1) \oplus \mathfrak{u}(1)$ symmetry where $a^\dagger_\pm$ and $b^\dagger_\pm$ have charges $(\pm1,0)$ and $(0,\pm 1)$ respectively.
While in the rational and trigonometric limits this follows from a global symmetry of the light-cone gauge-fixed Lagrangian, as shown in \appref{app:symmetries}, in the elliptic case there is no such global symmetry.
We can see this explicitly in the quadratic Lagrangian~\eqref{eq:Lag2elliptic}.
This hidden $\mathfrak{u}(1) \oplus \mathfrak{u}(1)$ symmetry only becomes manifest in the tree-level S-matrix once appropriate asymptotic states, with momentum-dependent coefficients, are identified.
It would be interesting to see if this hidden symmetry persists for higher-point scattering.

\section{Limits and relation to \texorpdfstring{$J \bar{T}$}{JTbar} and \texorpdfstring{$T \bar{T}$}{TTbar} deformations}
\label{sec:jttb}

In this section we analyse the S-matrix for different choices of the deformation parameters.
We restrict to the case in which $\AdS_3$ and $\Sp^3$ are deformed in the same way, i.e.~we set $\alpha_j=\beta_j$ for $j=1,2,3$.
In the rational ($\alpha_1=\alpha_2 =\alpha_3$) and trigonometric ($\alpha_1=\alpha_3$) limits, the S-matrix is related to the standard rational and trigonometric S-matrix for strings on $\AdS_3 \times \Sp^3$ by a $J\bar{T}$ deformation due to our non-standard choice of gauge~\cite{Frolov:2019xzi,Borsato:2023oru}.

\subsection{Rational limit}
\label{subsec:rationallimit}

In the limit $\alpha_j = \beta_j = 1$, or equivalently $\ga_1=\tg_1=\ga_3=\tg_3=1$ and $\tg_2=\ga_2=0$, the dispersion relations become
\begin{equation}
\tom_\pm(p) = \om_\pm(p) = \bar{\omega} \pm 1 ~, \qquad \bar{\omega} = \sqrt{p^2+1}~.
\end{equation}
The coefficients in the oscillator expansion \eqref{eq:o-W,-,eq:o-Zb} then simplify considerably to
\begin{equation} \label{eq:coeff-rational}
A_+ = B_- = C_+ = D_- = 0~, \qquad A_- = B_+ = C_- = D_+ =\frac{1}{\sqrt{\bar{\omega}}}~,
\end{equation}
which prompts the identification
\begin{equation}\label{eq:ident}
\ket{a^\dagger_+} = \ket{\overline{W}}~, \qquad \ket{a^\dagger_-} = \ket{W}~, \qquad \ket{b^\dagger_+} = \ket{\overline{Z}}~, \qquad \ket{b^\dagger_-} = \ket{Z}~.
\end{equation}
The scattering matrix elements \eqref{eq:Tmat-first,-,eq:Tmat-last} can be written as
\begin{align} \label{eq:Smat-rational}
\mathcal T \ket{a_{\mu_1}^\dagger a_{\mu_2}^\dagger} &= \left(-2 \bar{\mathcal A}_{\mu_1 \mu_2} - a ( p_1 \mu_2 - p_2 \mu_1) - \bar{\mathcal D} \right) \ket{a_{\mu_1}^\dagger a_{\mu_2}^\dagger}~, \\
\mathcal T \ket{a_{\mu_1}^\dagger a_{\mu_2}^\dagger} &= \left(+2 \bar{\mathcal A}_{\mu_1 \mu_2} - (a-1) ( p_1 \mu_2 - p_2 \mu_1) - \bar{\mathcal D} \right) \ket{b_{\mu_1}^\dagger b_{\mu_2}^\dagger}~, \\
\mathcal T \ket{a_{\mu_1}^\dagger b_{\mu_2}^\dagger} &= \left(+2 \bar{\mathcal G} - (a-1) p_1 \mu_2 + a p_2 \mu_1 - \bar{\mathcal D} \right) \ket{a_{\mu_1}^\dagger b_{\mu_2}^\dagger}~, \\
\mathcal T \ket{b_{\mu_1}^\dagger a_{\mu_2}^\dagger} &= \left(-2 \bar{\mathcal G} - a p_1 \mu_2 +(a-1) p_2 \mu_1 - \bar{\mathcal D} \right) \ket{b_{\mu_1}^\dagger a_{\mu_2}^\dagger}~,
\label{eq:Smatend}
\end{align}
with
\begin{equation} \begin{aligned}
\bar{\mathcal A}_{\mu_1 \mu_2} &= \frac{1}{4} \frac{(\mu_2 p_1 + \mu_1 p_2)^2 }{p_1 \bar{\omega}_2 - p_2 \bar{\omega}_1}~, \qquad
\bar{\mathcal G} = \frac{1}{4} (p_1 \bar{\omega}_2 + p_2 \bar{\omega}_1)~,
\\
\bar{\mathcal D} &= \left(a-\frac{1}{2}\right) (p_1 \bar{\omega}_2 - p_2 \bar{\omega_1})~, \qquad
\bar{\omega}_j = \bar{\omega}_{\mu_j}(p_j)~.
\end{aligned}
\end{equation}
The first terms in each expression above reproduce the usual tree-level S-matrix for $\AdS_3 \times \Sp^3$ in the standard light-cone gauge~\cite{Sundin:2013ypa}.
To better understand the additional terms, it is useful to use the relation between the excitations $(u,v,x,y)$ of the standard ``diagonal'' light-cone gauge fixing and $(U,V,X,Y)$ of our ``unilateral'' gauge fixing, discussed in \secref{sec:metricAdS3xS3}.
With the $\tau$-dependent rotation
\begin{equation}
W = w\,e^{i \tau}~, \qquad Z = z \, e^{i \tau}~,
\end{equation}
the quadratic Lagrangian \eqref{eq:L2ellipticAdS3} becomes, up to total derivatives, the same as the canonical one,
\begin{equation}
\Lag_2^r =|\dot{w}|^2 - |\acute{w}|^2 - |w|^2 + |\dot{z}|^2 - |\acute{z}|^2 - |z|^2~,
\end{equation}
describing two free gapped complex fields, with dispersion relation $\bar{\omega} = \sqrt{p^2+1}$, and equations of motion given by
\begin{equation}\begin{aligned}
\mathcal E_w &\coloneqq \ddot{w} - w'' +w =0~, \qquad  &\mathcal E_{\bar{w}} & \coloneqq \ddot{\bar{w}} - \bar{w}'' + \bar{w}=0~, \\ \mathcal E_z & \coloneqq \ddot{z} - z'' +z =0~, \qquad & \mathcal E_{\bar{z}} & \coloneqq \ddot{\bar{z}} - \bar{z}'' + \bar{z}=0~.
\end{aligned}\end{equation}

To write the quartic Lagrangian it is useful to look at the symmetries of $\Lag_2^r$ and derive the corresponding conserved currents.
The energy-momentum tensor gathers the conserved currents associated to translation invariance on the worldsheet of the string.
For the $\Lag_2^r$ the energy-momentum tensor is symmetric, with its components given by
\begin{equation} \label{eq:T-trigo}
\begin{aligned}
T_{\tau \tau}^r &= -|\dot{w}|^2-|\acute{w}|^2 - |w|^2-|\dot{z}|^2-|\acute{z}|^2 - |z|^2~, \\
T_{\sigma \sigma}^r &= -|\dot{w}|^2-|\acute{w}|^2 + |w|^2 -|\dot{z}|^2-|\acute{z}|^2 + |z|^2~, \\
T_{\tau \sigma}^r &= T_{ \sigma \tau}^r = - \dot{w} \acute{\bar{w}} - \acute{w} \dot{\bar{w}} - \dot{z} \acute{\bar{z}} - \acute{z} \dot{\bar{z}}~.
\end{aligned}
\end{equation}
In addition, there are also the conserved currents associated to the $\alg{u}(1)$ symmetries realised as $w \rightarrow e^{i \epsilon}w$ and $z \rightarrow e^{i \epsilon'}z$.
These read
\begin{equation}
J_\tau^w = w \dot{\bar{w}} - \dot{w} \bar{w} ~, \qquad J_\sigma^w = w \acute{\bar{w}} - \acute{w} \bar{w}~, \qquad J_\tau^z = z \dot{\bar{z}} - \dot{z} \bar{z} ~, \qquad J_\sigma^z = z \acute{\bar{z}} - \acute{z} \bar{z}~.
\label{eq:Jafterrotationundeformed}
\end{equation}
Using these we can then construct the following $T\bar{T}$ and $J \bar{T}$ operators
\begin{equation}
O_{T \bar{T}}^r = T_{\tau \tau}^r T_{\sigma \sigma}^r - T_{\tau \sigma}^r T_{\sigma \tau}^r, \qquad O_{J \bar{T}}^r = J_\tau T_{\sigma \sigma}^r - J_\sigma T_{\tau \sigma}^r~.
\label{eq:TTbarJTbaroperundeformed}
\end{equation}
There are two different $J\bar{T}$ operators, one where the current is taken to be $J=J^w$ and the other where it is taken to be $J=J^z$.
Note that the energy-momentum tensor \eqref{eq:T-trigo} is related to the energy-momentum \eqref{eq:Tmunu} after rotating the fields through the relations
\begin{equation}
T_{\tau \tau} \rightarrow T_{\tau \tau}^r - i J_\tau^w - i J_\tau^z~, \qquad T_{\tau \sigma} \rightarrow T_{\tau \sigma}^r~, \qquad T_{\sigma \tau} \rightarrow T_{\sigma \tau}^r - i J_\sigma^w - i J_\sigma^z~,\qquad T_{\sigma \sigma} \rightarrow T_{\sigma \sigma}^r ~,
\end{equation}
which in particular implies that upon rotation
\begin{equation}
-\big(a-\frac{1}{2} \big)O_{T \bar{T}} \rightarrow -\big( a- \frac{1}{2} \big) O_{T \bar{T}}^r + i a O_{J^w \bar{T}}^r + i (a-1) O_{J^z \bar{T}}^r - \frac{i}{2} O_{J^w \bar{T}}^r + \frac{i}{2} O_{J^z \bar{T}}^r~.
\end{equation}
The quartic Lagrangian can then be written
\begin{equation} \begin{aligned}
\Lag_4^r &= -2 |w|^2 |\acute{w}|^2+2 |z|^2 |\acute{z}|^2 - |w|^2 |\dot{z}|^2 + |\dot{w}|^2 |z|^2 - |w|^2 |\acute{z}|^2 + |\acute{w}|^2 |z|^2 \\
&\qquad -\big(a-\frac{1}{2}\big) O_{T \bar{T}}^r + i a O_{J^w \bar{T}}^r + i (a-1) O_{J^z \bar{T}}^r+ \dots
\end{aligned}
\end{equation}
where the ellipses denote terms that explicitly depend on worldsheet time $\tau$, but vanish on the equations of motion.
The term proportional to $(a-\frac{1}{2})$ corresponds to a $T\bar{T}$ deformation, which is usually encountered when fixing uniform light-cone gauge~\cite{Frolov:2019nrr,Baggio:2018gct}.
Due to the non-standard light-cone gauge fixing, we encounter two new terms: one proportional to $a$ corresponding to a $J \bar{T}$ deformation with current $J=J^w$, and one proportional to $(a-1)$ corresponding to a $J \bar{T}$ deformation with current $J=J^z$.
These additional terms also appear in the S-matrix \eqref{eq:Smat-rational,-,eq:Smatend}.
Indeed, the effect of a $T\bar{T}$ deformation on the S-matrix of a 2-d integrable model is known, and given by dressing by a phase
\begin{equation}
\mathcal S(p_1,p_2) \rightarrow e^{i (a-\frac12)(p_1 \bar{\omega}_2 - p_2 \bar{\omega}_1)} \mathcal S(p_1,p_2)~,
\end{equation}
which in the tree-level S-matrix gives the term proportional to $(a-\frac{1}{2})$ in eq.~\eqref{eq:Smat-rational,-,eq:Smatend}.
The effect of the $J \bar{T}$ deformation is a twist involving momentum and the $\alg{u}(1)$ charge of the excitations, which produces the middle terms on the right-hand side of \eqref{eq:Smat-rational,-,eq:Smatend}.
For a discussion on the connections between different gauge choices and $J\bar{T}$ deformations see~\cite{Frolov:2019xzi,Borsato:2023oru}.

\subsection{Trigonometric limit}
\label{subsec:trigonometriclimit}

If we choose the deformation parameters
\begin{equation}
\al_1 = \al_3 = \be_1 = \be_3 = \frac{1}{1+\kappa^2}~, \qquad \al_2=\be_2=1~,
\end{equation}
or equivalently
\begin{equation}
\tg_1=\ga_1=1+\kappa^2, \qquad \tg_2=\ga_2=0, \qquad \tg_3=\ga_3=1-\kappa^2~,
\end{equation}
the dispersion relation becomes
\begin{equation}
\omega_\pm^a(p) = \omega_\pm^b(p) = \bar{\omega}_\pm \pm 1~, \qquad \bar{\omega}_\pm=\sqrt{p^2 +(1+\kappa^2)^2} \mp \kappa^2~.
\end{equation}
As expected $\bar{\omega}_\pm$ is the dispersion relation of the unilateral YB deformation.
\unskip\footnote{Note that $\bar{\omega}_+$ and $\bar{\omega}_-$ are interchanged with respect to some literature, in particular~\cite{Seibold:2021lju} and~\cite{Bocconcello:2020qkt}.}
The coefficients in the mode expansion again simplify, with
\begin{equation}
A_+ = B_- = C_+ = D_- =0~, \qquad A_- = B_+ = C_- = D_+ = \frac{1}{(p^2 + (1+\kappa^2)^2)^{\frac{1}{4}}}~,
\end{equation}
prompting the same identification as in the rational case between fields and oscillators~\eqref{eq:ident}.
The scattering elements can then be written as
\begin{align} \label{eq:Smat-trigo}
\mathcal T \ket{a_{\mu_1}^\dagger(p_1) a_{\mu_2}^\dagger(p_2)} &= (- 2\bar{\mathcal A}_{\mu_1 \mu_2} - a (p_1 \mu_2 - p_2 \mu_1)- \bar{\mathcal D}_{\mu_1 \mu_2}) \ket{a_{\mu_1}^\dagger(p_1) a_{\mu_2}^\dagger(p_2)} ~, \\
\mathcal T \ket{b_{\mu_1}^\dagger(p_1) b_{\mu_2}^\dagger(p_2)} &= (+ 2 \bar{\mathcal A}_{\mu_1 \mu_2} -(a-1)(p_1 \mu_2 - p_2 \mu_1)- \bar{\mathcal D}_{\mu_1 \mu_2}) \ket{b_{\mu_1}^\dagger(p_1) b_{\mu_2}^\dagger(p_2)}~, \\
\mathcal T \ket{a_{\mu_1}^\dagger(p_1) b_{\mu_2}^\dagger(p_2)} &= (+2 \bar{\mathcal G}_{\mu_1 \mu_2} -(a-1) p_1 \mu_2 + a p_2 \mu_1-\bar{\mathcal D}_{\mu_1 \mu_2} ) \ket{a_{\mu_1}^\dagger(p_1) b_{\mu_2}^\dagger(p_2)}~, \\
\mathcal T \ket{b_{\mu_1}^\dagger(p_1) a_{\mu_2}^\dagger(p_2)} &= (-2 \bar{\mathcal G}_{\mu_1 \mu_2} -a p_1 \mu_2 + (a-1) p_2 \mu_1-\bar{\mathcal D}_{\mu_1 \mu_2}) \ket{b_{\mu_1}^\dagger(p_1) a_{\mu_2}^\dagger(p_2)}~,
\end{align}
with
\begin{equation} \begin{aligned}
\bar{\mathcal A}_{\mu_1 \mu_2} &= \frac{1}{4} \frac{(\mu_2 p_1 + \mu_1 p_2)^2 + 2 \kappa^2 (\mu_2 \bar{\omega}_1 + \mu_1 \bar{\omega}_2)^2-2 \kappa^2 (\mu_1 \mu_2 \bar{\omega}_1\bar{\omega}_2 +1)(\mu_1 \bar{\omega}_1 + \mu_2 \bar{\omega}_2)}{p_1 \bar{\omega}_2 - p_2 \bar{\omega}_1}~, \\
\bar{\mathcal G}_{\mu_1 \mu_2} &= \frac{1}{4} (p_1 \bar{\omega}_2 + p_2 \bar{\omega}_1)~, \qquad
\bar{\mathcal D}_{\mu_1 \mu_2} = \left(a-\frac{1}{2}\right) (p_1 \bar{\omega}_2 - p_2 \bar{\omega_1})~, \qquad
\bar{\omega}_j = \bar{\omega}_{\mu_j}(p_j)~.
\end{aligned}
\end{equation}
The scattering amplitudes $\bar{\mathcal A}$, $\bar{\mathcal B}$ and $\bar{\mathcal D}$ are precisely those corresponding to the unilateral YB deformation, where we again recall that compared to some literature $\mu \rightarrow -\mu$.
As expected, they can be directly obtained starting from \eqref{eq:metric-YB} and performing the standard light-cone gauge fixing in $t$ and $\varphi$.
As in the rational case, the additional terms correspond to $J \bar{T}$ deformations.

This relation can again be seen at the level of the light-cone gauge-fixed Lagrangians.
Implementing the same $\tau$-dependent rotation as in the rational case, namely
\begin{equation}
W = w e^{i \tau}~, \qquad Z= z e^{i \tau}~,
\end{equation}
gives the quadratic Lagrangian
\unskip\footnote{
Note that we can make a further time-dependent rotation such that the quadratic Lagrangian \eqref{eq:L2-trigo} canonically describes two gapped complex fields with mass $1+\kappa^2$.}
\begin{equation} \label{eq:L2-trigo}
\Lag_2^r = |\dot{w}|^2 - |\acute{w}|^2 - (1+2 \kappa^2)|w|^2+i \kappa^2(w\dot{\bar{w}}-\dot{w}\bar{w}) + |\dot{z}|^2 - |\acute{z}|^2 - (1+2 \kappa^2)|z|^2+i \kappa^2(z\dot{\bar{z}}-\dot{z}\bar{z})~,
\end{equation}
which indeed corresponds to the quadratic Lagrangian of the unilateral YB deformation.
The corresponding equations of motion are given by
\begin{equation}
\begin{aligned}
&\mathcal E_w \coloneqq \ddot{w} - w''+2i\kappa^2 \dot{w} +(1+2 \kappa^2)w =0~, \qquad &&\mathcal E_{\bar{w}} \coloneqq \ddot{\bar{w}} - \bar{w}'' -2i\kappa^2 \dot{\bar{w}}+ (1+2 \kappa^2)\bar{w}=0~, \\
& \mathcal E_z \coloneqq \ddot{z} - z''+2i\kappa^2 \dot{z} +(1+2 \kappa^2)z =0~, \qquad &&\mathcal E_{\bar{z}} \coloneqq \ddot{\bar{z}} - \bar{z}'' -2i\kappa^2 \dot{\bar{z}}+ (1+2 \kappa^2)\bar{z}=0~.
\end{aligned}
\end{equation}
Computing the conserved charges for this theory gives
\begin{equation}
\begin{aligned}
&T_{\tau\tau}^r=-|\dot{w}|^2-|\acute{w}|^2-(1+2\kappa^2)|w|^2-|\dot{z}|^2-|\acute{z}|^2-(1+2\kappa^2)|z|^2~,\\
&T_{\sigma\sigma}^r=-|\dot{w}|^2-|\acute{w}|^2+(1+2\kappa^2)|w|^2-i\kappa^2(w\dot{\bar{w}}-\dot{w}\bar{w})\\
&\hspace{1cm}-|\dot{z}|^2-|\acute{z}|^2+(1+2\kappa^2)|z|^2-i\kappa^2(z\dot{\bar{z}}-\dot{z}\bar{z})~,\\
&T_{\tau\sigma}^r=-\dot{w}\acute{\bar{w}}-\acute{w}\dot{\bar{w}}-i\kappa^2(w\acute{\bar{w}}-\acute{w}\bar{w})-\dot{z}\acute{\bar{z}}-\acute{z}\dot{\bar{z}}-i\kappa^2(z\acute{\bar{z}}-\acute{z}\bar{z})~,\\
&T_{\sigma\tau}^r=-\dot{w}\acute{\bar{w}}-\acute{w}\dot{\bar{w}}-\dot{z}\acute{\bar{z}}-\acute{z}\dot{\bar{z}}~,
\end{aligned}
\end{equation}
for the energy-momentum tensor and
\begin{equation}
\begin{aligned}
&J_\tau^w = w \dot{\bar{w}} - \dot{w} \bar{w}-2i\kappa^2|w|^2~, && J_\tau^z = z \dot{\bar{z}} - \dot{z} \bar{z}-2i\kappa^2|z|^2~, \\
&J_\sigma^w = w \acute{\bar{w}} - \acute{w} \bar{w}~, && J_\sigma^z = z \acute{\bar{z}} - \acute{z} \bar{z}~,
\end{aligned}
\label{eq:Jafterrotationtrig}
\end{equation}
for the conserved currents associated with the $\alg{u}(1)$ symmetries.
We can now define the $ T\bar{T} $ and $ J\bar{T} $ operators as in eq.~\eqref{eq:TTbarJTbaroperundeformed}.
With this, the quartic Lagrangian $ \Lag_4^r $ can again be written in such a way that the terms proportional to $(a-\frac{1}{2})$ are given by the $ T\bar{T} $ operator, the terms proportional to $ a $ by a $ J\bar{T} $ deformation associated to the current in $ \AdS_3 $ and the terms proportional to $(a-1)$ by a $ J\bar{T} $ operator associated to the current in $\Sp^3$.
Explicitly we can write, up to total derivatives and terms proportional to the equations of motion,
\begin{equation}
\Lag_4^r = \check{\Lag}_4^r + \hat{\Lag}_4^r + \tilde{\Lag}_4^r -\big(a-\frac{1}{2}\big)O_{T\bar{T}}^r+iaO_{J^w\bar{T}}^r+i(a-1)O_{J^z\bar{T}}^r + \dots ~,
\end{equation}
with
\begin{align}
\check{\Lag}_4^r &= -2\kappa^2|w|^2|\dot{w}|^2-2(1+\kappa^2)|w|^2|\acute{w}|^2-\kappa^2|w|^4+\frac{3}{2}i\kappa^2|w|^2(w\dot{\bar{w}}-\dot{w}\bar{w})\nonumber \\
&\quad +\frac{1}{2}\kappa^2\left(w(w+i\dot{w})(\dot{\bar{w}}^2-\acute{\bar{w}}^2)+\bar{w}(\bar{w}-i\dot{\bar{w}})(\dot{w}^2-\acute{w}^2)\right)~, \\
\hat{\Lag}_4^r &= - \left.\check{\Lag}_4^r \right|_{w \rightarrow z}~, \\
\tilde{\Lag}_4^r &= (1+2\kappa^2)\left(|z|^2|\dot{w}|^2+|z|^2|\acute{w}|^2-|w|^2|\dot{z}|^2-|w|^2|\acute{z}|^2\right)\nonumber\\
&\quad +\frac{i\kappa^2}{2}\left((w \dot{\bar{w}}-\dot{w}\bar{w})(|\dot{z}|^2+|\acute{z}|^2+(1+2 \kappa^2) |z|^2)-(z \dot{\bar{z}}-\dot{z}\bar{z})(|\dot{w}|^2+|\acute{w}|^2+(1+2 \kappa^2) |w|^2)\right) \nonumber \\
&\quad -\frac{i\kappa^2}{2}\left((w \acute{\bar{w}}-\acute{w}\bar{w})(\dot{z}\acute{\bar{z}}+\acute{z}\dot{\bar{z}})-(z \acute{\bar{z}}-\acute{z}\bar{z})(\dot{w}\acute{\bar{w}}+\acute{w}\dot{\bar{w}})\right)~.
\end{align}
The three terms $\check{\Lag}_4^r + \hat{\Lag}_4^r + \tilde{\Lag}_4^r$ constitute the quartic Lagrangian obtained in the standard light-cone gauge fixing with $a=\frac{1}{2}$.

\section{Factorisation and comparison to an exact elliptic S-matrix}
\label{sec:exact}

For the $\AdS_3 \times \Sp^3$ superstring and its YB deformation, the exact S-matrix computed in the standard light-cone gauge (involving global time $t$ in $\AdS_3$ and the ``diagonal'' angle $\varphi$ in $\Sp^3$) can be bootstrapped on symmetry grounds and is given by the tensor product of a factorised S-matrix (up to dressing factors) 
\begin{equation} \label{eq:S-factorised}
\mathcal S = S \otimes S~.
\end{equation}
The factorised S-matrix $S$ is invariant under the centrally-extended algebra $\mathcal A =[\alg{su}(1|1)_\ind{L} \oplus \alg{su}(1|1)_\ind{R}]_{\mathrm{c.e.}}$.
This factorisation also applies to the asymptotic states, which are given by the tensor products
\begin{equation}
\ket{b_+^\dagger} = \ket{\phi_+ \otimes \phi_+}~, \qquad \ket{b_-^\dagger} = \ket{\phi_- \otimes \phi_-}~, \qquad \ket{a^\dagger_+} = \ket{\psi_+ \otimes \psi_+}~, \qquad \ket{a_-^\dagger} = \ket{\psi_- \otimes \psi_-}~,
\end{equation}
where $(\phi_+|\psi_+)$ and $(\phi_-|\psi_-)$ transform in fundamental (``left'' and ``right'') representations of $\mathcal A$.
The other tensor products describe the gapped fermions.
Recalling that we only consider the gapped sector of $\AdS_3 \times \Sp^3 \times \To^4$, $S$ further decomposes into the direct sum of four $4 \times 4$ blocks describing the left-left, left-right, right-left and right-right scattering.

In this section we would like to understand if we can find a similar factorisation for the elliptically-deformed model that is also compatible with known results in the rational and trigonometric limits. 
The perturbative computation in \secref{sec:tlsmat} fixes the diagonal tree-level entries of the $4 \times 4$ blocks. 
However, in the rational and trigonometric limits, the tree-level factorised S-matrix is not purely diagonal. 
To directly determine the off-diagonal entries in the elliptic case, we would need to compute the tree-level scattering of the gapped fermions. 
We will instead follow an indirect approach, which is to assume that the tree-level factorised S-matrix solves the classical Yang-Baxter equation, and then attempt to solve for the non-vanishing unknown entries. 
As we will see, we find that there is no such solution indicating that either factorisation, the direct sum structure or integrability is broken upon including fermions. 

In the rational and trigonometric cases, the $4 \times 4$ blocks are known to be of 6-vertex type, and are contained in the 6vB R-matrix of~\cite{deLeeuw:2021ufg}.
A deformation of the rational case built from four 8-vertex blocks, the 8vB R-matrix, and conjectured to correspond to an elliptic integrable deformation has been constructed in~\cite{deLeeuw:2021ufg}.
As the name suggests, this S-matrix has eight non-vanishing entries.
In the left-left or right-right sectors we have
\begin{equation}
\begin{aligned}
S \ket{\phi_\pm(p_1) \phi_\pm(p_2)} &= r_{1,\pm \pm} \ket{\phi_\pm(p_1) \phi_\pm(p_2)} + r_{8,\pm \pm} \ket{\psi_\pm (p_1) \psi_\pm(p_2)}~, \\
S \ket{\phi_\pm(p_1) \psi_\pm(p_2)} &= r_{2,\pm \pm} \ket{\phi_\pm(p_1) \psi_\pm(p_2)} + r_{6,\pm \pm} \ket{\psi_\pm (p_1) \phi_\pm(p_2)}~, \\
S \ket{\psi_\pm(p_1) \phi_\pm(p_2)} &= r_{3,\pm \pm} \ket{\psi_\pm(p_1) \phi_\pm(p_2)} + r_{5,\pm \pm} \ket{\phi_\pm (p_1) \psi_\pm(p_2)}~, \\
S \ket{\psi_\pm(p_1) \psi_\pm(p_2)} &= r_{4,\pm \pm}\ket{\psi_\pm(p_1) \psi_\pm(p_2)} + r_{7,\pm \pm}\ket{\psi_\pm (p_1) \psi_\pm(p_2)}~, \\
\end{aligned}
\end{equation}
while for the left-right and right-left sectors we have
\begin{equation}
\begin{aligned}
S \ket{\phi_\pm(p_1) \phi_\mp(p_2)} &= r_{1,\pm \mp} \ket{\phi_\pm(p_1) \phi_\mp(p_2)} + r_{8,\pm \mp} \ket{\psi_\pm (p_1) \psi_\mp(p_2)}~, \\
S \ket{\phi_\pm(p_1) \psi_\mp(p_2)} &= r_{2,\pm \mp} \ket{\phi_\pm(p_1) \psi_\mp(p_2)} + r_{6,\pm \mp} \ket{\psi_\pm (p_1) \phi_\mp(p_2)}~, \\
S \ket{\psi_\pm(p_1) \phi_\pm(p_2)} &= r_{3,\pm \mp} \ket{\psi_\pm(p_1) \phi_\mp(p_2)} + r_{5,\pm \mp} \ket{\phi_\pm (p_1) \psi_\mp(p_2)}~, \\
S \ket{\psi_\pm(p_1) \psi_\pm(p_2)} &= r_{4,\pm \mp}\ket{\psi_\pm(p_1) \psi_\mp(p_2)} + r_{7,\pm \mp}\ket{\psi_\pm (p_1) \psi_\mp(p_2)}~.
\end{aligned}
\end{equation}
The case $r_{7,\pm \pm} = r_{8,\pm \pm} = r_{5, \pm \mp} = r_{6, \pm \mp} =0$ corresponds to the 6-vertex limit.
Again expanding the factorised S-matrix as
\begin{equation}
S = 1+iT + \dots ~,
\end{equation}
we denote the entries of the tree-level S-matrix $T$ by $\tilde{r}_{j,\mu_1,\mu_2}$.
Expanding~\eqref{eq:S-factorised} we find
\begin{equation} \begin{aligned}
\mathcal T \ket{a_{\mu_1}^\dagger a^\dagger_{\mu_2}} &= 2 \tilde{r}_{4,\mu_1 \mu_2} \ket{a_{\mu_1}^\dagger a^\dagger_{\mu_2}} \qquad &\Rightarrow\qquad \tilde{r}_{4, \mu_1 \mu_2} &= - \mathcal A_{\mu_1 \mu_2}^{aa} - \frac{1}{2} \mathcal B_{\mu_1 \mu_2}^{aa} - \frac{1}{2} \mathcal D_{\mu_1 \mu_2}^{aa}~, \\
\mathcal T \ket{b_{\mu_1}^\dagger b^\dagger_{\mu_2}} &= 2 \tilde{r}_{1,\mu_1 \mu_2} \ket{b_{\mu_1}^\dagger b^\dagger_{\mu_2}} \qquad &\Rightarrow\qquad \tilde{r}_{1, \mu_1 \mu_2} &= + \mathcal A_{\mu_1 \mu_2}^{bb} + \frac{1}{2} \mathcal B_{\mu_1 \mu_2}^{bb} - \frac{1}{2} \mathcal D_{\mu_1 \mu_2}^{bb}~, \\
\mathcal T \ket{a_{\mu_1}^\dagger b^\dagger_{\mu_2}} &= 2 \tilde{r}_{3, \mu_1 \mu_2} \ket{a_{\mu_1}^\dagger b^\dagger_{\mu_2}} \qquad &\Rightarrow\qquad \tilde{r}_{3, \mu_1 \mu_2} &= + \mathcal G_{\mu_1 \mu_2}^{ab} - \frac{1}{2} \mathcal D_{\mu_1 \mu_2}^{ab}~, \\
\mathcal T \ket{b_{\mu_1}^\dagger a^\dagger_{\mu_2}} &= 2 \tilde{r}_{2, \mu_1 \mu_2} \ket{b_{\mu_1}^\dagger a^\dagger_{\mu_2}} \qquad &\Rightarrow\qquad \tilde{r}_{2, \mu_1 \mu_2} &= - \mathcal G_{\mu_1 \mu_2}^{ba} - \frac{1}{2} \mathcal D_{\mu_1 \mu_2}^{ba}~.
\end{aligned}\label{eq:Smat}
\end{equation}
Therefore, at tree-level we cannot distinguish between 6-vertex and 8-vertex since the additional matrix entries $r_5, r_6, r_7, r_8$ do not contribute at this order in the boson-boson scattering.

The 6-vertex 6vB and 8-vertex 8vB R-matrices, which describe deformations of the undeformed $4\times 4$ blocks, both satisfy the free-fermion condition \cite{DeLeeuw:2020ahx}
\begin{equation}
r_1 r_4 - r_2 r_3 = r_5 r_6 - r_7 r_8~.
\label{eq:FFcondition}
\end{equation}
In particular, all known $ \AdS_3 \times\Sp^3$ S-matrices in the literature and their integrable deformations satisfy this condition.
\unskip\footnote{The presence of the free-fermion condition in S-matrices related to AdS backgrounds had previously been noticed in~\cite{Mitev:2012vt} and~\cite{Hoare:2014kma}.}
Expanding the free-fermion condition imposes the following constraint on the tree-level scattering elements
\begin{equation}
\tilde{r}_{1, \mu_1 \mu_2} + \tilde{r}_{4,\mu_1 \mu_2} = \tilde{r}_{2,\mu_1 \mu_2} + \tilde{r}_{3,\mu_1 \mu_2}~.
\end{equation}
From the expressions in \eqref{eq:Smat} it follows that the free-fermion condition is satisfied for the elliptically deformed S-matrix computed in \secref{sec:tlsmat} provided that $\alpha_j=\beta_j$ for $j=1,2,3$.
In other words, $\AdS_3$ and $\Sp^3$ should be deformed in the same way.

Let us now address the question of whether our elliptic model is of 6-vertex or 8-vertex type, or neither, by directly solving the classical YB equation.
In order to do this we assume that for $ \mu_1=\mu_2=+1 $, $ \tilde{r}_{j} $ with $ j=1,2,3,4 $ are given by eq.~\eqref{eq:Smat}.
We then substitute them, together with unspecified $ \tilde{r}_{k} $ for $ k=5,6,7,8 $, into the classical YB equation \eqref{eq:cYBE}.
Finally, we try to solve for the unknown functions $ \tilde{r}_{k} $ such that the classical YB equation is satisfied.

For the 6-vertex case, namely the case with $ \tilde{r}_{7} =\tilde{r}_{8}=0$, we quickly find an inconsistency.
When solving the equations, the expression for $ \tilde{r}_{5}(p_1,p_3)\tilde{r}_{6}(p_1,p_3)$ is equal to another expression that depends on $ p_2 $.
We checked that this is the case both algebraically and numerically, and also observed that the inconsistency disappears in the rational and trigonometric limits.
The calculation for the 8-vertex case ($ \tilde{r}_{7} \neq0$ and $\tilde{r}_{8}\neq0$ ) is more involved and needs to be done numerically, but it ultimately leads to the same type of inconsistencies.
These results indicate that if there exists an integrable string sigma model admitting the tree-level S-matrix \eqref{eq:Tmat-first,-,eq:Tmat-last} as its bosonic truncation, then the exact S-matrix does not take the factorised form \eqref{eq:S-factorised} with $S$ of 6-vertex or 8-vertex type as considered in~\cite{deLeeuw:2020ahe,deLeeuw:2021ufg}.
In turn this suggests that either the full S-matrix cannot be factorised or the factorised S-matrix itself cannot be written as a direct sum of $ 4\times 4 $ blocks.

A further possibility is that integrability is broken.
The 8vB R-matrix of~\cite{deLeeuw:2020ahe,deLeeuw:2021ufg} is not a deformation of the $q$-deformed S-matrix in~\cite{Hoare:2014oua,Seibold:2021lju}, which describes the trigonometric deformation of $\AdS_3 \times \Sp^3 \times \To^4$, whereas the elliptically-deformed sigma model is a deformation of the trigonometric one.
In order for the 8vB R-matrix to have the $q$-deformation as a limit an extra twist is necessary.
However, the introduction of this twist breaks integrability.
This opens up the possibility that including fermions breaks integrability in a controlled way with the elliptic deformation studied here related to the twisted 8vB R-matrix.

In order to include fermions we need to embed the elliptic deformation of the bosonic $\AdS_3 \times \Sp^3$ sigma model in type II supergravity.
In the following section we construct such embeddings, however we postpone the inclusion of fermions in the S-matrix computation to future work.

\section{Supergravity solutions}
\label{sec:sugra}

In \secref{sec:lcgft} we considered the light-cone gauge fixing of the bosonic model~\eqref{eq:Lag-sigma-model}.
In order for the light-cone gauge-fixed theory to make sense beyond tree-level, we would like to find a 10-d type II supergravity embedding of the bosonic background~\eqref{eq:ellipticmetricAdS3xS3}.
We start by completing the 6-d elliptic deformation of $\AdS_3 \times \Sp^3$ with a 4-torus, i.e.~adding four free compact bosons to~\eqref{eq:Lag-sigma-model}.

Since we know that the bosonic model is integrable, we would like to find a deformed supergravity background that describes an integrable string sigma model.
We take the undeformed background to be $\AdS_3 \times \Sp^3 \times \To^4$ supported by pure R-R 3-form flux, which has $\grp{PSU}(1,1|2) \times \grp{PSU}(1,1|2)$ global symmetry, preserving half the maximal 10-d supersymmetry, i.e.~16 supercharges.
The elliptic deformation preserves the left-acting $\grp{SL}(2,\Real) \times \grp{SU}(2)$ global symmetry,
Therefore, ideally our deformed background would also preserve one copy of $\grp{PSU}(1,1|2)$.

Such backgrounds for trigonometric deformations were studied in detail in~\cite{Hoare:2022asa}, and our construction is motivated by that paper.
In particular, we look for a type IIB supergravity background that takes the following form
\begin{equation}\begin{gathered}
ds^2 = g_{\mu\nu}(\fields^\rho) d\fields^\mu d\fields^\nu + \sum_{r=6}^9 d\fields_r d\fields_r ~, \qquad
H_3 = 0~, \qquad \Phi = 0 ~,
\\
F_1 = 0 ~, \qquad
F_3 = F_3^{(4)}(\fields^\rho) ~, \qquad
F_5 = \sum_{i=1}^3 F_3^{(i)}(\fields^\rho) \wedge J_2^{(i)} ~,
\end{gathered}\end{equation}
where $H_3$ is the NS-NS flux, $F_{1,3,5}$ are the R-R fluxes and $\Phi$ is the dilaton.
\unskip\footnote{We have set the constant dilaton $\Phi$ to zero since it is straightforward to restore its dependence.}
The index $\mu,\nu,\rho,\dots = 0,\ldots,5$ runs over the $\AdS_3$ and $\Sp^3$ directions, i.e.~$\{\Psi^\rho\} = \{T,U,V,\Phi,X,Y\}$, while $r,\ldots = 6,\ldots,9$ labels the 4-torus directions.
The 2-forms
\begin{equation}
J_2^{(1)} = d\fields_6 \wedge d\fields_7 - d\fields_8 \wedge d\fields_9 ~, \qquad
J_2^{(2)} = d\fields_6 \wedge d\fields_8 + d\fields_7 \wedge d\fields_9 ~, \qquad
J_2^{(3)} = d\fields_6 \wedge d\fields_9 - d\fields_7 \wedge d\fields_8 ~,
\end{equation}
are three orthogonal self-dual 2-forms on the 4-torus.
Finally, the 3-form and 5-form R-R fluxes are parametrised in terms of four closed 3-forms, $F_3^{(i)}$, $i = 1,\ldots, 4$, which we take to only depend on the $\AdS_3$ and $\Sp^3$ directions.
We furthermore assume that they are self-dual $\star_6 F_3^{(i)} = F_3^{(i)}$, implying that $(d \star_6 F_3^{(i)} = 0)=|F_3^{(i)}|^2 = 0$, and that they are orthogonal, i.e.~$F_3^{(i)} \cdot F_3^{(j)} = 0$ for $i \neq j$.
Under these assumptions the type IIB supergravity equations simplify to
\begin{equation}\label{eq:typeiib}
R_{\mu\nu} - \frac{1}{4}\big(\sum_{i=1}^4 (F_3^{(i)})_{\mu\rho\sigma}(F_3^{(i)})_{\nu}{}^{\rho\sigma}\big) = 0 ~.
\end{equation}

Introducing the $(\grp{SL}(2,\Real) \times \grp{SU}(2))$-invariant vielbein for the elliptically-deformed $\AdS_3 \times \Sp^3$
\begin{equation}\begin{aligned}
e^0 & = \frac{\sqrt{\alpha_2}}2 \Tr(g^{-1} d g \cdot L_2) = \sqrt{\alpha_2} (-\cosh2U\cosh2V \, dT - \sinh 2V \, dU) ~,
\\
e^1 & = \frac{\sqrt{\alpha_1}}2 \Tr(g^{-1} d g \cdot L_1) = \sqrt{\alpha_1}(-\sinh2U \, dT + dV)
\\
e^2 & = \frac{\sqrt{\alpha_3}}2 \Tr(g^{-1} d g \cdot L_3) = \sqrt{\alpha_3}(\cosh2U\sinh2V \, dT + \cosh2V \, dU) ~,
\\
e^3 & = \frac{\sqrt{\beta_2}}2 \Tr(g^{-1} d g \cdot J_2) = \sqrt{\beta_2}( - \cos2X\cos2Y \, d\Phi + \sin 2Y \, dX) ~,
\\
e^4 & = \frac{\sqrt{\beta_1}}2 \Tr(g^{-1} d g \cdot J_1) = \sqrt{\beta_1}(\sin2X \, d\Phi - dY) ~,
\\
e^5 & = \frac{\sqrt{\beta_3}}2 \Tr(g^{-1} d g \cdot J_3) = \sqrt{\beta_3}(-\cos2X\sin2Y \, d\Phi - \cos 2Y \, dX) ~,
\end{aligned}\end{equation}
where the generators $L_i$ and $J_i$, and the group element $g$ are defined in eqs.~\eqref{eq:lmat,-,eq:param}, we have that
\begin{equation}
g_{\mu\nu}(\fields^\rho) d\fields^\mu d\fields^\nu = -(e^0)^2 + (e^1)^2 + (e^2)^2 + (e^3)^2 + (e^4)^2 + (e^5)^2 ~.
\end{equation}
Let us introduce the following basis of 3-form fluxes
\begin{equation}\begin{gathered}\label{eq:fluxbasis}
f_3^{(1)} = d(e^1 \wedge e^4) = 2 (
-\frac{\sqrt{\alpha_1}}{\sqrt{\alpha_2\alpha_3}} e^0 \wedge e^2 \wedge e^4
+ \frac{\sqrt{\beta_1}}{\sqrt{\beta_2\beta_3}} e^1 \wedge e^3 \wedge e^5) ~,
\\
f_3^{(2)} = d(e^0 \wedge e^3) = 2 (
-\frac{\sqrt{\alpha_2}}{\sqrt{\alpha_1\alpha_3}} e^1 \wedge e^2 \wedge e^3
- \frac{\sqrt{\beta_2}}{\sqrt{\beta_1\beta_3}} e^0 \wedge e^4 \wedge e^5) ~,
\\
f_3^{(3)} = - d(e^2 \wedge e^5) = 2 (
-\frac{\sqrt{\alpha_3}}{\sqrt{\alpha_1\alpha_2}} e^0 \wedge e^1 \wedge e^5
+ \frac{\sqrt{\beta_3}}{\sqrt{\beta_1\beta_2}} e^2 \wedge e^3 \wedge e^4) ~,
\\
f_3^{(4)} = 2 (e^0 \wedge e^1 \wedge e^2 + e^3 \wedge e^4 \wedge e^5) ~.
\end{gathered}\end{equation}
Requiring these to be self-dual implies that
\begin{equation}\label{eq:betaalpha}
\beta_1 = \alpha_1 ~, \qquad \beta_2 = \alpha_2 ~, \qquad \beta_3 = \alpha_3 ~.
\end{equation}
Moreover, contracting the equation of motion~\eqref{eq:typeiib} with $g^{\mu\nu}$, we see that this implies the Ricci scalar should vanish.
This is indeed the case if we take $\alpha_i$ and $\beta_i$ to be related as in eq.~\eqref{eq:betaalpha}.
From this point on we will assume that eq.~\eqref{eq:betaalpha} holds, i.e. $\AdS_3$ and $\Sp^3$ are deformed in the same way.

We now take the following ansatz for our four 3-form fluxes
\begin{equation}\label{eq:fluxes}
F_3^{(i)} = \frac{\sqrt{\alpha_2\alpha_3}}{\sqrt{\alpha_1}} \mathbf{x}^{(1)}_i f_3^{(1)}
+ \frac{\sqrt{\alpha_1\alpha_3}}{\sqrt{\alpha_2}} \mathbf{x}^{(2)}_i f_3^{(2)}
+ \frac{\sqrt{\alpha_1\alpha_2}}{\sqrt{\alpha_3}} \mathbf{x}^{(3)}_i f_3^{(3)}
+ \mathbf{x}^{(4)}_i f_3^{(4)} ~.
\end{equation}
Substituting into the equation of motion~\eqref{eq:typeiib}, we find that it is solved if we set
\begin{equation}\begin{aligned}\label{eq:solution}
||\mathbf{x}^{(1)}||^2 & = \frac{\alpha_2 + \alpha_3 - \alpha_1}{\alpha_2 \alpha_3} - ||\mathbf{x}^{(4)}||^2 ~,
& \qquad
\mathbf{x}^{(1)} \cdot \mathbf{x}^{(4)} & = - \mathbf{x}^{(2)} \cdot \mathbf{x}^{(3)} ~,
\\
||\mathbf{x}^{(2)}||^2 & = \frac{\alpha_2 - \alpha_1 - \alpha_3}{\alpha_1 \alpha_3} + ||\mathbf{x}^{(4)}||^2 ~,
& \qquad
\mathbf{x}^{(2)} \cdot \mathbf{x}^{(4)} & = + \mathbf{x}^{(1)} \cdot \mathbf{x}^{(3)} ~,
\\
||\mathbf{x}^{(3)}||^2 & = \frac{\alpha_2 + \alpha_1 - \alpha_3}{\alpha_1 \alpha_2} - ||\mathbf{x}^{(4)}||^2 ~,
& \qquad
\mathbf{x}^{(3)} \cdot \mathbf{x}^{(4)} & = - \mathbf{x}^{(1)} \cdot \mathbf{x}^{(2)} ~.
\end{aligned}
\end{equation}
Note that since the equation of motion~\eqref{eq:typeiib} is invariant under rotations of the four 3-form fluxes $F_3^{(i)}$ amongst themselves, this solution is $\grp{SO}(4)$ invariant, i.e.~if we have one solution we can freely apply an $\grp{SO}(4)$ transformation to find a new solution.
This corresponds to the freedom to apply TsT transformations in the 4-torus directions.
Furthermore, if we had allowed for a non-zero NS-NS flux satisfying the same properties as $F_3^{(i)}$, then the type IIB supergravity equations simplify to
\begin{equation}
R_{\mu\nu} - \frac{1}{4} H_3{}_{\mu\rho\sigma} H_{3}{}_\nu{}^{\rho\sigma} - \frac{1}{4}\big(\sum_{i=1}^4 (F_3^{(i)})_{\mu\rho\sigma}(F_3^{(i)})_{\nu}{}^{\rho\sigma}\big) = 0 ~,
\end{equation}
which is invariant under rotations of the five 3-form fluxes $H_3$ and $F_3^{(i)}$ amongst themselves.
Therefore, starting from the solution~\eqref{eq:fluxes} with~\eqref{eq:solution}, we can apply $\grp{SO}(5)$ transformations to find new solutions with non-vanishing NS-NS flux.
The simplicity of these solutions, with constant dilaton and homogeneous fluxes, suggests they are good candidates to describe an integrable string sigma model.

Given the form of the solution in~\eqref{eq:solution}, it is important to ask if there exist real solutions.
Indeed, if we look at the undeformed limit $\alpha_1 = \alpha_2 = \alpha_3 = \Tstr$, we find
\begin{equation}
||\mathbf{x}^{(1)}||^2 = - ||\mathbf{x}^{(2)}||^2 = ||\mathbf{x}^{(3)}||^2 = \Tstr^{-1} - ||\mathbf{x}^{(4)}||^2 ~.
\end{equation}
To ensure that the solution is real, we immediately see that we need to take
\begin{equation}
||\mathbf{x}^{(4)}||^2 = \Tstr^{-1} ~, \qquad
\mathbf{x}^{(1)} = \mathbf{x}^{(2)} = \mathbf{x}^{(3)} = 0 ~.
\end{equation}
This corresponds to the familiar $\AdS_3 \times \Sp^3 \times \To^4$ background supported by pure R-R flux with 16 supercharges with $\Tstr$ playing the role of the string tension.

The trigonometric deformation discussed in~\cite{Hoare:2022asa} corresponds to taking $\alpha_1 = \alpha_3 = \frac{1}{1+\kappa^2} \Tstr$ and $\alpha_2 = \Tstr$, for which
\begin{equation}
||\mathbf{x}^{(1)}||^2 = ||\mathbf{x}^{(3)}||^2 = (1+\kappa^2) \Tstr^{-1} - ||\mathbf{x}^{(4)}||^2 ~, \qquad
||\mathbf{x}^{(2)}||^2 = (-1+\kappa^2)(1+\kappa^2)\Tstr^{-1} + ||\mathbf{x}^{(4)}||^2 ~.
\end{equation}
In this case a real solution only exists if we take $(1-\kappa^2)(1+\kappa^2)\Tstr^{-1} \leq ||\mathbf{x}^{(4)}||^2 \leq (1+\kappa^2) \Tstr^{-1}$, which requires $\kappa^2 \geq 0$.
Moreover, we can preserve 8 supercharges if we choose $||\mathbf{x}^{(4)}||^2 = (1+\kappa^2)\Tstr^{-1}$, in which case we exactly recover the integrable background of~\cite{Hoare:2022asa}.
This suggests that if there is a choice of fluxes~\eqref{eq:fluxes} supporting the 6-d elliptic deformation of $\AdS_3 \times \Sp^3$ preserving 8 supercharges then this may require an additional restriction on top of~\eqref{eq:solution}.

\section{Conclusions}
\label{sec:conc}

In this paper we computed the bosonic tree-level S-matrix for an elliptic deformation of the $\grp{SL}(2;\mathbb R) \times \grp{SU}(2)$ sigma model, which, after adding the 4-torus directions, can be thought of as an elliptic deformation of the $\AdS_3 \times \Sp^3 \times \To^4$ string.
This deformation only preserves two $\alg{u}(1)$ isometries, which we used to fix uniform light-cone gauge.
While the quadratic and quartic light-cone gauge Lagrangians do not have any manifest symmetries, after defining appropriate asymptotic states the tree-level S-matrix only exhibits diagonal scattering.
The classical YB equation is then trivially satisfied, in agreement with the integrability of the elliptic model.
We checked that in the rational and trigonometric deformation limit our results are in agreement with the known perturbative S-matrix for strings propagating in $\AdS_3 \times \Sp^3$~\cite{Sundin:2013ypa} and YB-deformed $\AdS_3 \times \Sp^3$~\cite{Seibold:2021lju} respectively.
Since the two $\alg{u}(1)$ isometries used to fix uniform light-cone gauge do not correspond to the usual global time in $\AdS_3$ and great circle in $\Sp^3$, our results are related to the standard perturbative S-matrices by a $J\bar{T}$ deformation~\cite{Frolov:2019xzi,Borsato:2023oru}.
To pave the way to analysing elliptic deformations of the $\AdS_3 \times \Sp^3 \times \To^4$ superstring we also constructed an explicit embedding of the elliptically-deformed metric in type IIB supergravity.

An interesting open problem is the formulation of an exact elliptically-deformed S-matrix that solves the quantum YB equation and reproduces our tree-level results in the large tension limit.
Such an exact S-matrix can in general be bootstrapped from the symmetries of the light-cone gauge-fixed theory, up to overall dressing factors whose construction relies on additional physical input~\cite{Beisert:2005tm,Arutyunov:2006ak}.
In particular, for superstrings on $\AdS_3 \times \Sp^3 \times \To^4$ quantised in the standard uniform light-cone gauge, the original $\alg{psu}(1,1|2) \oplus \alg{psu}(1,1|2)$ algebra is broken to a centrally-extended $\left[\alg{su}(1|1) \oplus \alg{su}(1|1)\right]_{\mathrm{c.e.}}^{\oplus 2}$ algebra~\cite{Borsato:2014exa}.
For the trigonometric deformation, only the four $\alg{u}(1)$s are manifestly realised in the sigma model.
Fortunately, there is a hidden quantum-group symmetry~\cite{Delduc:2013fga,Delduc:2014kha} which again makes it possible to bootstrap the S-matrix~\cite{Hoare:2014oua}.
In the elliptic case, a careful study of the symmetries of the deformed sigma model and its gauge-fixed version remains to be carried out.

Another approach to finding an exact S-matrix is through a classification of solutions to the quantum YB equation, as initiated in~\cite{deLeeuw:2020ahe}.
For strings on $\AdS_3 \times \Sp^3 \times \To^4$ in standard light-cone gauge, the full $64 \times 64$ S-matrix describing the scattering in the gapped sector can be written as a restricted tensor product of two factorised $16 \times 16$ S-matrices.
This factorised S-matrix can further be decomposed into four $4 \times 4$ blocks $S_\ind{LL}$, $S_\ind{LR}$, $S_\ind{RL}$ and $S_\ind{RR}$.
In~\cite{deLeeuw:2021ufg}, assuming that $S_\ind{LL}$ and $S_\ind{RR}$ are both of 6-vertex or 8-vertex type, the space of possible integrable $16 \times 16$ S-matrices was explored.
The standard $\AdS_3$ string S-matrix and its trigonometric deformation lie in the 6-vertex case.
A new elliptic S-matrix built out of 8-vertex blocks was also found.
However, our perturbative results do not appear to be compatible with the large tension expansion of this 8-vertex S-matrix.
Since the classification in~\cite{deLeeuw:2021ufg} is complete, this implies that at least one of the three assumptions made there does not apply to our model after including fermions. 
Namely, either the full S-matrix does not factorise, the factorised S-matrix cannot be written as a direct sum of $4\times 4$ blocks, or integrability is broken. 
To gain a better understanding of this it would be insightful to compute the tree-level S-matrix for the supergravity embedding of \secref{sec:sugra} to see if integrability is still present once the fermions are included. 

Another interesting possibility is to add an additional B-field to the construction.
Strings on $\AdS_3$ can be supported by a mixture of R-R and NS-NS fluxes, preserving their integrability~\cite{Cagnazzo:2012se}.
The symmetry algebra of the theory, hence also the structure of the S-matrix, is the same across the moduli space~\cite{Lloyd:2014bsa}.
However, the representation parameters used to match the exact S-matrix with the perturbative results~\cite{Hoare:2013pma,Hoare:2013lja} depend on the amount of NS-NS flux.
The dressing phases also need to be modified, and are yet to be fully determined, although see~\cite{OhlssonSax:2023qrk} for recent progress in this direction.
In the case of the trigonometric deformation, it is known that it is possible to add a B-field while preserving integrability~\cite{Delduc:2014uaa,Delduc:2018xug} and the embedding in type II supergravity was found in~\cite{Hoare:2022asa}.
However, how the exact S-matrix is affected by this additional contribution is not yet clear.
Indeed, the tree-level S-matrix for a three-parameter deformation of the $\AdS_3 \times \Sp^3 \times \To^4$ string, which includes the trigonometric case with B-field, was constructed in~\cite{Bocconcello:2020qkt}, but has not yet been matched to an exact S-matrix.
For elliptic deformations it is not known if it is possible to add a B-field to the construction while preserving integrability.

Finally, other deformations of the $\grp{SL}(2;\Real) \times \grp{SU}(2)$ sigma model can be considered, in particular the $\lambda$-deformation~\cite{Balog:1993es,Sfetsos:2013wia,Hollowood:2014rla,Hollowood:2014qma}, which is related to the trigonometric deformation considered in this paper through Poisson-Lie duality~\cite{Klimcik:1995jn,Klimcik:1995ux,Vicedo:2015pna,Hoare:2015gda,Hoare:2017ukq}.
The $\lambda$-deformation can be generalised to $\Integer_4$ permutation supercosets~\cite{Hoare:2022vnw}, giving rise to a model preserving two manifest $\alg{u}(1)$ symmetries and 8 supersymmetries.
An explicit supergravity embedding for this supersymmetric model has been proposed in~\cite{Itsios:2023kma}.
Bosonic scattering for the $\lambda$-deformed $\grp{SL}(2;\Real) \times \grp{SU}(2)$ sigma model was analysed in~\cite{Georgiou:2022fow}, and it would be interesting to include fermions.
It is an open question if there is a notion of Poisson-Lie duality and an associated $\lambda$-deformation for the elliptic model.

\subsection*{Acknowledgements}
ALR and FS thank the participants of the Workshop “Integrability in Low Supersymmetry Theories” in Filicudi, Italy for stimulating discussions.
We also would like to thank Sergey Frolov for pointing out the connection between the choice of light-cone gauge fixing and $JT$ transformations, Alessandro Torrielli, Davide Polvara and Sylvain Lacroix for discussions, and Sibylle Driezen and Riccardo Borsato for collaboration on a related project.
The work of BH and ALR was supported by a UKRI Future Leaders Fellowship (grant number MR/T018909/1).
The work of FS was supported by the European Union Horizon 2020 research and innovation programme under the Marie Sklodowska-Curie grant agreement number 101027251.

\appendix

\section{Symmetries}
\label{app:symmetries}

In this appendix we analyse in more detail the symmetries of the action \eqref{eq:Lag-sigma-model} and its metric \eqref{eq:ellipticmetricAdS3xS3}.
Since the $\AdS_3$ part of the action can be obtained by analytic continuation, we focus on $\Sp^3$.

\paragraph{Undeformed symmetries before light-cone gauge fixing.}
The undeformed action is invariant under left- and right-multiplications by global $\grp{SU}(2)$ elements.
Infinitesimally, the left transformations act on the fields as
\begin{align}
\delta_\ind{L}^1 &= \{ \Phi \rightarrow \Phi + \epsilon \sin(2 \Phi) \tan(2 X), X \rightarrow X+\epsilon \cos(2 \Phi), Y \rightarrow Y+ \epsilon \sin(2 \Phi) \sec (2 X)\}~, \\
\delta_\ind{L}^2 &= \{ \Phi \rightarrow \Phi + \epsilon \cos(2 \Phi) \tan (2 X), X \rightarrow X - \epsilon \sin(2 \Phi), Y \rightarrow Y+ \epsilon \cos(2 \Phi) \sec(2 X) \}~, \\
\delta_\ind{L}^3 &= \{ \Phi \rightarrow \Phi+\epsilon, X \rightarrow X, Y \rightarrow Y \}~.
\end{align}
The undeformed metric \eqref{eq:metric-undef} is invariant under these transformations (up to $\mathcal O(\epsilon^2)$ terms), which can be shown to explicitly satisfy an $\alg{su}(2)$ algebra.
For the right symmetry the transformation rules are similar,
\begin{align}
\delta_\ind{R}^1 &= \{ \Phi \rightarrow \Phi+ \epsilon \sin(2 Y) \sec (2 X), X \rightarrow X+\epsilon \cos(2 Y), Y \rightarrow Y + \epsilon \sin(2 Y) \tan(2 X)\}~, \\
\delta_\ind{R}^2 &= \{ \Phi \rightarrow \Phi+ \epsilon \cos(2 Y) \sec(2 X), X \rightarrow X - \epsilon \sin(2 Y), Y \rightarrow Y + \epsilon \cos(2 Y) \tan (2 X) \}~, \\
\delta_\ind{R}^3 &= \{ \Phi \rightarrow \Phi , X \rightarrow X,Y \rightarrow Y+\epsilon\}~.
\end{align}
The associated conserved charges are $Q = \int d \sigma \mathcal Q$ with
\begin{align}
\mathcal Q^1_\ind{L} &= -p_X \cos(2 \Phi) - \sec(2 X) \sin(2 \Phi) (p_Y + p_\Phi \sin(2 X))~, \\
\mathcal Q^2_\ind{L} &= + p_X \sin(2 \Phi) - \sec(2 X) \cos(2 \Phi)(p_Y + p_\Phi \sin(2 X)) ~, \\
\mathcal Q^3_\ind{L} &= - p_\Phi ~,
\end{align}
and
\begin{align}
\mathcal Q^1_\ind{R} &= -p_X \cos(2 Y) - \sec(2 X) \sin(2 Y) (p_\Phi + p_Y \sin(2 X))~, \\
\mathcal Q^2_\ind{R} &= + p_X \sin(2 Y) - \sec(2 X) \cos(2 Y)(p_\Phi + p_Y \sin(2 X)) ~, \\
\mathcal Q^3_\ind{R} &= -p_Y~.
\end{align}
Assuming canonically normalised fields, $\{\field(\sigma),P_{\field}(\sigma')\}=i \delta(\sigma-\sigma')$ at equal time $\tau$, we have
\begin{align}
\{\mathcal Q_\ind{L}^\alpha(\sigma), \mathcal Q_\ind{R}^\beta(\sigma')\} &=0 ~, \\
\{\mathcal Q_\ind{L}^\alpha(\sigma), \mathcal Q_\ind{L}^\beta(\sigma')\} &= 2i \epsilon^{\alpha \beta \gamma} \mathcal Q_\ind{L}^\gamma (\sigma) \delta(\sigma-\sigma')~, \\
\{\mathcal Q_\ind{R}^\alpha(\sigma), Q_\ind{R}^\beta(\sigma') \} &= 2 i \epsilon^{\alpha \beta \gamma} \mathcal Q_\ind{R}^\gamma(\sigma) \delta(\sigma-\sigma')~.
\end{align}
As expected, we recover the $\alg{su}(2)_\ind{L} \oplus \alg{su}(2)_\ind{R}$ algebra.

\paragraph{Undeformed symmetries after light-cone gauge fixing.}
Upon light-cone gauge fixing
\begin{equation}
p_\Phi = a p_+ + p_-~, \qquad p_- = 1~, \qquad p_+ = - H~,
\end{equation}
the conserved charges associated to right multiplications do not depend explicitly on $X^+=\tau$, hence Poisson-commute with the worldsheet Hamiltonian, as follows from the conservation law
\begin{equation}
0=\frac{d Q}{d \tau} = \frac{\partial Q}{\partial \tau} + \{ Q, H \} ~.
\end{equation}
Therefore the $\alg{su}(2)_\ind{R}$ algebra survives the light-cone gauge fixing.
For the left charges, $\mathcal Q_\ind{L}^3 = -p_\Phi = a H -1$ trivially commutes with the worldsheet Hamiltonian $H$.
The other two charges $\mathcal Q_\ind{L}^{1}$ and $\mathcal Q_\ind{L}^2$ depend explicitly on $\Phi$, hence are broken by the light-cone gauge fixing.

\paragraph{Deformed symmetries.}
In the presence of the deforming operator $\mathcal O$ with generic deformation parameters, only the left $\alg{su}(2)_\ind{L}$ symmetry remains before light-cone gauge fixing.
The infinitesimal transformation rules leaving the action invariant are the same as in the undeformed case above, and the conserved charges take the same form when written in terms of the fields and their conjugate momenta.
As before, with the exception of $Q_\ind{L}^3$, which is just a constant plus a term proportional to the Hamiltonian, the left charges do not survive the light-cone gauge fixing.
The only non-trivial symmetries in the light-cone gauge-fixed action arise from right multiplications.
For the right symmetry, applying the above transformation rules we find that
\begin{equation}
\delta \mathcal L^1_\ind{R} \sim \be_1 - \be_2~, \qquad \delta \mathcal L^2_\ind{R} \sim \be_1 - \be_3~, \qquad \delta \mathcal L^3_\ind{R} \sim \be_2 - \be_3~.
\end{equation}
When all the deformation parameters are equal we recover the undeformed case described above with an $\alg{su}(2)_\ind{R}$ symmetry.
If two deformation parameters are equal, then there is a surviving $\alg{u}(1)_\ind{R}$ symmetry.
In particular, for the trigonometric deformation with $\be_1=\be_3$, the charge $Q \equiv -Q_\ind{R}^2$ is conserved.
Its expansion in terms of oscillators gives a quadratic contribution
\begin{equation}
Q_2 = b_+^\dagger b_+ -b_-^\dagger b_-~,
\end{equation}
indicating that the excitations generated by $b_+^\dagger$ and $b_-^\dagger$ carry charge $+1$ and $-1$ respectively.
For generic deformation parameters (all different from each other), the symmetry is completely broken.
We thus naively do not expect any remaining $\alg{u}(1)$ symmetry in the light-cone gauge-fixed theory and its S-matrix.
However, our results obtained in \secref{sec:tlsmat} suggest that the S-matrix still possesses a hidden $\alg{u}(1)$ symmetry once appropriate asymptotic states, with momentum-dependent coefficients, are identified.

\begin{bibtex}[\jobname]

@article{Sundin:2013ypa,
author = "Sundin, Per and Wulff, Linus",
title = "{Worldsheet scattering in AdS(3)/CFT(2)}",
eprint = "1302.5349",
archivePrefix = "arXiv",
primaryClass = "hep-th",
reportNumber = "MIFPA-13-08",
doi = "10.1007/JHEP07(2013)007",
journal = "JHEP",
volume = "07",
pages = "007",
year = "2013"
}

@article{Vicedo:2015pna,
author = "Vicedo, Benoit",
title = "{Deformed integrable \ensuremath{\sigma}-models, classical R-matrices and classical exchange algebra on Drinfel\textquoteright{}d doubles}",
eprint = "1504.06303",
archivePrefix = "arXiv",
primaryClass = "hep-th",
doi = "10.1088/1751-8113/48/35/355203",
journal = "J. Phys. A",
volume = "48",
number = "35",
pages = "355203",
year = "2015"
}

@article{Klimcik:1995ux,
author = "Klim\v{c}\'{i}k, C. and \v{S}evera, P.",
title = "{Dual nonAbelian duality and the Drinfeld double}",
eprint = "hep-th/9502122",
archivePrefix = "arXiv",
reportNumber = "CERN-TH-95-39, CERN-TH-95-039",
doi = "10.1016/0370-2693(95)00451-P",
journal = "Phys. Lett. B",
volume = "351",
pages = "455--462",
year = "1995"
}

@article{Klimcik:1995jn,
author = "Klim\v{c}\'{i}k, C.",
editor = "Gava, E. and Narain, K. S. and Vafa, C.",
title = "{Poisson-Lie T duality}",
eprint = "hep-th/9509095",
archivePrefix = "arXiv",
reportNumber = "CERN-TH-95-248",
doi = "10.1016/0920-5632(96)00013-8",
journal = "Nucl. Phys. B Proc. Suppl.",
volume = "46",
pages = "116--121",
year = "1996"
}

@article{Hoare:2015gda,
author = "Hoare, B. and Tseytlin, A. A.",
title = "{On integrable deformations of superstring sigma models related to $AdS_n \times S^n$ supercosets}",
eprint = "1504.07213",
archivePrefix = "arXiv",
primaryClass = "hep-th",
reportNumber = "IMPERIAL-TP-AT-2015-02, HU-EP-15-21",
doi = "10.1016/j.nuclphysb.2015.06.001",
journal = "Nucl. Phys. B",
volume = "897",
pages = "448--478",
year = "2015"
}

@article{Hoare:2017ukq,
author = "Hoare, Ben and Seibold, Fiona K.",
title = "{Poisson-Lie duals of the $\eta$ deformed symmetric space sigma model}",
eprint = "1709.01448",
archivePrefix = "arXiv",
primaryClass = "hep-th",
doi = "10.1007/JHEP11(2017)014",
journal = "JHEP",
volume = "11",
pages = "014",
year = "2017"
}

@article{Georgiou:2022fow,
author = "Georgiou, George and Sfetsos, Konstantinos",
title = "{Scattering in integrable pp-wave backgrounds: S-matrix and absence of particle production}",
eprint = "2208.01072",
archivePrefix = "arXiv",
primaryClass = "hep-th",
reportNumber = "CERN-TH-2022-129",
doi = "10.1016/j.nuclphysb.2023.116096",
journal = "Nucl. Phys. B",
volume = "987",
pages = "116096",
year = "2023"
}

@article{Sfetsos:2014cea,
author = "Sfetsos, Konstantinos and Thompson, Daniel C.",
title = "{Spacetimes for $\lambda$-deformations}",
eprint = "1410.1886",
archivePrefix = "arXiv",
primaryClass = "hep-th",
doi = "10.1007/JHEP12(2014)164",
journal = "JHEP",
volume = "12",
pages = "164",
year = "2014"
}

@article{Balog:1993es,
author = "Balog, J. and Forgacs, P. and Horvath, Z. and Palla, L.",
editor = "Lust, D. and Weigt, G.",
title = "{A New family of SU(2) symmetric integrable sigma models}",
eprint = "hep-th/9307030",
archivePrefix = "arXiv",
reportNumber = "ITP-502-BUDAPEST",
doi = "10.1016/0370-2693(94)90213-5",
journal = "Phys. Lett. B",
volume = "324",
pages = "403--408",
year = "1994"
}

@article{Sfetsos:2013wia,
author = "Sfetsos, Konstadinos",
title = "{Integrable interpolations: From exact CFTs to non-Abelian T-duals}",
eprint = "1312.4560",
archivePrefix = "arXiv",
primaryClass = "hep-th",
reportNumber = "DMUS-MP-13-23, DMUS--MP--13-23",
doi = "10.1016/j.nuclphysb.2014.01.004",
journal = "Nucl. Phys. B",
volume = "880",
pages = "225--246",
year = "2014"
}

@article{Hoare:2022vnw,
author = "Hoare, Ben and Levine, Nat and Seibold, Fiona K.",
title = "{Bi-\ensuremath{\eta} and bi-\ensuremath{\lambda} deformations of \ensuremath{\Integer}$_{4}$ permutation supercosets}",
eprint = "2212.08625",
archivePrefix = "arXiv",
primaryClass = "hep-th",
reportNumber = "Imperial-TP-FS-2022-03",
doi = "10.1007/JHEP04(2023)024",
journal = "JHEP",
volume = "04",
pages = "024",
year = "2023"
}

@article{Itsios:2023kma,
    author = "Itsios, Georgios and Sfetsos, Konstantinos and Siampos, Konstantinos",
    title = "{Supersymmetric backgrounds from \ensuremath{\lambda}-deformations}",
    eprint = "2310.17700",
    archivePrefix = "arXiv",
    primaryClass = "hep-th",
    reportNumber = "HU-EP-23/57",
    doi = "10.1007/JHEP01(2024)084",
    journal = "JHEP",
    volume = "01",
    pages = "084",
    year = "2024"
}

@article{Hollowood:2014qma,
author = "Hollowood, Timothy J. and Miramontes, J. Luis and Schmidtt, David M.",
title = "{An Integrable Deformation of the $AdS_5 \times S^5$ Superstring}",
eprint = "1409.1538",
archivePrefix = "arXiv",
primaryClass = "hep-th",
doi = "10.1088/1751-8113/47/49/495402",
journal = "J. Phys. A",
volume = "47",
number = "49",
pages = "495402",
year = "2014"
}

@article{Hollowood:2014rla,
author = "Hollowood, Timothy J. and Miramontes, J. Luis and Schmidtt, David M.",
title = "{Integrable Deformations of Strings on Symmetric Spaces}",
eprint = "1407.2840",
archivePrefix = "arXiv",
primaryClass = "hep-th",
doi = "10.1007/JHEP11(2014)009",
journal = "JHEP",
volume = "11",
pages = "009",
year = "2014"
}

@article{Klose:2006zd,
author = "Klose, Thomas and McLoughlin, Tristan and Roiban, Radu and Zarembo, Konstantin",
title = "{Worldsheet scattering in $AdS_5 \times S^5$}",
eprint = "hep-th/0611169",
archivePrefix = "arXiv",
reportNumber = "ITEP-TH-61-06, UUITP-15-06",
doi = "10.1088/1126-6708/2007/03/094",
journal = "JHEP",
volume = "03",
pages = "094",
year = "2007"
}

@article{Arutyunov:2005hd,
author = "Arutyunov, Gleb and Frolov, Sergey",
title = "{Uniform light-cone gauge for strings in $AdS_5 \times S^5$: Solving SU(1|1) sector}",
eprint = "hep-th/0510208",
archivePrefix = "arXiv",
reportNumber = "ITP-UU-05-47, SPIN-05-32, AEI-2005-160",
doi = "10.1088/1126-6708/2006/01/055",
journal = "JHEP",
volume = "01",
pages = "055",
year = "2006"
}

@article{Frolov:2019xzi,
author = "Frolov, Sergey",
title = "{$T{\overline T}$, $\widetilde JJ$, $JT$ and $\widetilde JT$ deformations}",
eprint = "1907.12117",
archivePrefix = "arXiv",
primaryClass = "hep-th",
doi = "10.1088/1751-8121/ab581b",
journal = "J. Phys. A",
volume = "53",
number = "2",
pages = "025401",
year = "2020"
}

@article{Hoare:2022asa,
author = "Hoare, Ben and Seibold, Fiona K. and Tseytlin, Arkady A.",
title = "{Integrable supersymmetric deformations of AdS$_{3}$\texttimes{} S$^{3}$\texttimes{} T$^{4}$}",
eprint = "2206.12347",
archivePrefix = "arXiv",
primaryClass = "hep-th",
reportNumber = "Imperial-TP-AT-2022-02",
doi = "10.1007/JHEP09(2022)018",
journal = "JHEP",
volume = "09",
pages = "018",
year = "2022"
}

@article{Hoare:2013pma,
author = "Hoare, B. and Tseytlin, A. A.",
title = "{On string theory on $AdS_3 \times S^3 \times T^4$ with mixed 3-form flux: tree-level S-matrix}",
eprint = "1303.1037",
archivePrefix = "arXiv",
primaryClass = "hep-th",
reportNumber = "IMPERIAL-TP-AT-2013-01, HU-EP-13-10",
doi = "10.1016/j.nuclphysb.2013.05.005",
journal = "Nucl. Phys. B",
volume = "873",
pages = "682--727",
year = "2013"
}

@article{Cagnazzo:2012se,
author = "Cagnazzo, A. and Zarembo, K.",
title = "{B-field in AdS(3)/CFT(2) Correspondence and Integrability}",
eprint = "1209.4049",
archivePrefix = "arXiv",
primaryClass = "hep-th",
reportNumber = "NORDITA-2012-67, UUITP-24-12",
doi = "10.1007/JHEP11(2012)133",
journal = "JHEP",
volume = "11",
pages = "133",
year = "2012",
note = "Erratum: \texttt{JHEP 04, 003 (2013)}"
}

@article{Delduc:2014uaa,
author = "Delduc, Francois and Magro, Marc and Vicedo, Benoit",
title = "{Integrable double deformation of the principal chiral model}",
eprint = "1410.8066",
archivePrefix = "arXiv",
primaryClass = "hep-th",
doi = "10.1016/j.nuclphysb.2014.12.018",
journal = "Nucl. Phys. B",
volume = "891",
pages = "312--321",
year = "2015"
}

@article{Delduc:2018xug,
author = "Delduc, F. and Hoare, B. and Kameyama, T. and Lacroix, S. and Magro, M.",
title = "{Three-parameter integrable deformation of $\Integer_4$ permutation supercosets}",
eprint = "1811.00453",
archivePrefix = "arXiv",
primaryClass = "hep-th",
reportNumber = "ZMP-HH/18-22",
doi = "10.1007/JHEP01(2019)109",
journal = "JHEP",
volume = "01",
pages = "109",
year = "2019"
}

@article{Lloyd:2014bsa,
author = "Lloyd, Thomas and Ohlsson Sax, Olof and Sfondrini, Alessandro and Stefa\'nski, Jr., Bogdan",
title = "{The complete worldsheet S matrix of superstrings on AdS$_3 \times$ S$^3 \times$ T$^4$ with mixed three-form flux}",
eprint = "1410.0866",
archivePrefix = "arXiv",
primaryClass = "hep-th",
reportNumber = "IMPERIAL-TP-OOS-2014-04, HU-MATHEMATIK-2014-21, HU-EP-14-34",
doi = "10.1016/j.nuclphysb.2014.12.019",
journal = "Nucl. Phys. B",
volume = "891",
pages = "570--612",
year = "2015"
}

@article{Delduc:2013fga,
author = "Delduc, Francois and Magro, Marc and Vicedo, Benoit",
title = "{On classical $q$-deformations of integrable sigma-models}",
eprint = "1308.3581",
archivePrefix = "arXiv",
primaryClass = "hep-th",
doi = "10.1007/JHEP11(2013)192",
journal = "JHEP",
volume = "11",
pages = "192",
year = "2013"
}

@article{Delduc:2014kha,
author = "Delduc, Francois and Magro, Marc and Vicedo, Benoit",
title = "{Derivation of the action and symmetries of the $q$-deformed $AdS_{5} \times S^{5}$ superstring}",
eprint = "1406.6286",
archivePrefix = "arXiv",
primaryClass = "hep-th",
doi = "10.1007/JHEP10(2014)132",
journal = "JHEP",
volume = "10",
pages = "132",
year = "2014"
}

@article{Arutyunov:2006ak,
author = "Arutyunov, Gleb and Frolov, Sergey and Plefka, Jan and Zamaklar, Marija",
title = "{The Off-shell Symmetry Algebra of the Light-cone $AdS_5 \times S^5$ Superstring}",
eprint = "hep-th/0609157",
archivePrefix = "arXiv",
reportNumber = "AEI-2006-071, HU-EP-06-31, ITP-UU-06-39, SPIN-06-33, TCDMATH-06-13",
doi = "10.1088/1751-8113/40/13/018",
journal = "J. Phys. A",
volume = "40",
pages = "3583--3606",
year = "2007"
}

@article{Beisert:2005tm,
author = "Beisert, Niklas",
title = "{The SU(2|2) dynamic S-matrix}",
eprint = "hep-th/0511082",
archivePrefix = "arXiv",
reportNumber = "PUTP-2181, NSF-KITP-05-92",
doi = "10.4310/ATMP.2008.v12.n5.a1",
journal = "Adv. Theor. Math. Phys.",
volume = "12",
pages = "945--979",
year = "2008"
}

@article{Borsato:2014exa,
author = "Borsato, Riccardo and Ohlsson Sax, Olof and Sfondrini, Alessandro and Stefanski, Bogdan",
title = "{Towards the All-Loop Worldsheet S Matrix for $AdS_3\times S^3\times T^4$}",
eprint = "1403.4543",
archivePrefix = "arXiv",
primaryClass = "hep-th",
reportNumber = "IMPERIAL-TP-OOS-2014-01, HU-MATHEMATIK-2014-05, HU-EP-14-12, SPIN-14-11, ITP-UU-14-10",
doi = "10.1103/PhysRevLett.113.131601",
journal = "Phys. Rev. Lett.",
volume = "113",
number = "13",
pages = "131601",
year = "2014"
}

@article{Arutyunov:2014jfa,
author = "Arutyunov, Gleb and van Tongeren, Stijn J.",
title = "{Double Wick rotating Green-Schwarz strings}",
eprint = "1412.5137",
archivePrefix = "arXiv",
primaryClass = "hep-th",
doi = "10.1007/JHEP05(2015)027",
journal = "JHEP",
volume = "05",
pages = "027",
year = "2015"
}

@article{Arutyunov:2009ga,
author = "Arutyunov, Gleb and Frolov, Sergey",
title = "{Foundations of the AdS$_{5} \times S^{5}$ Superstring. Part I}",
eprint = "0901.4937",
archivePrefix = "arXiv",
primaryClass = "hep-th",
reportNumber = "ITP-UU-09-05, SPIN-09-05, TCD-MATH-09-06, HMI-09-03",
doi = "10.1088/1751-8113/42/25/254003",
journal = "J. Phys. A",
volume = "42",
pages = "254003",
year = "2009"
}

@article{Seibold:2021lju,
author = "Seibold, Fiona K. and van Tongeren, Stijn J. and Zimmermann, Yannik",
title = "{On quantum deformations of AdS$_{3}$ \texttimes{} S$^{3}$ \texttimes{} T$^{4}$ and mirror duality}",
eprint = "2107.02564",
archivePrefix = "arXiv",
primaryClass = "hep-th",
reportNumber = "Imperial-TP-FS-2021-01, HU-EP-21/19",
doi = "10.1007/JHEP09(2021)110",
journal = "JHEP",
volume = "09",
pages = "110",
year = "2021"
}

@article{Bocconcello:2020qkt,
author = "Bocconcello, Marco and Masuda, Isari and Seibold, Fiona K. and Sfondrini, Alessandro",
title = "{S matrix for a three-parameter integrable deformation of AdS$_{3}$ \texttimes{} S$^{3}$ strings}",
eprint = "2008.07603",
archivePrefix = "arXiv",
primaryClass = "hep-th",
doi = "10.1007/JHEP11(2020)022",
journal = "JHEP",
volume = "11",
pages = "022",
year = "2020"
}

@article{deLeeuw:2020ahe,
author = "de Leeuw, Marius and Paletta, Chiara and Pribytok, Anton and Retore, Ana L. and Ryan, Paul",
title = "{Classifying Nearest-Neighbor Interactions and Deformations of AdS}",
eprint = "2003.04332",
archivePrefix = "arXiv",
primaryClass = "hep-th",
doi = "10.1103/PhysRevLett.125.031604",
journal = "Phys. Rev. Lett.",
volume = "125",
number = "3",
pages = "031604",
year = "2020"
}

@article{deLeeuw:2021ufg,
author = "de Leeuw, Marius and Pribytok, Anton and Retore, Ana L. and Ryan, Paul",
title = "{Integrable deformations of AdS/CFT}",
eprint = "2109.00017",
archivePrefix = "arXiv",
primaryClass = "hep-th",
doi = "10.1007/JHEP05(2022)012",
journal = "JHEP",
volume = "05",
pages = "012",
year = "2022"
}

@article{DeLeeuw:2020ahx,
author = "De Leeuw, Marius and Paletta, Chiara and Pribytok, Anton and Retore, Ana L. and Torrielli, Alessandro",
title = "{Free Fermions, vertex Hamiltonians, and lower-dimensional AdS/CFT}",
eprint = "2011.08217",
archivePrefix = "arXiv",
primaryClass = "hep-th",
reportNumber = "DMUS-MP-20/09; TCDMATH-20-14, DMUS-MP-20/09",
doi = "10.1007/JHEP02(2021)191",
journal = "JHEP",
volume = "02",
pages = "191",
year = "2021"
}

@article{Lacroix:2023qlz,
author = "Lacroix, Sylvain and Wallberg, Anders",
title = "{An elliptic integrable deformation of the Principal Chiral Model}",
eprint = "2311.09301",
archivePrefix = "arXiv",
primaryClass = "hep-th",
reportNumber = "CERN-TH-2023-205",
month = "11",
year = "2023"
}

@article{Hoare:2014oua,
author = "Hoare, Ben",
title = "{Towards a two-parameter q-deformation of AdS$_3 \times S^3 \times M^4$ superstrings}",
eprint = "1411.1266",
archivePrefix = "arXiv",
primaryClass = "hep-th",
reportNumber = "HU-EP-14-44",
doi = "10.1016/j.nuclphysb.2014.12.012",
journal = "Nucl. Phys. B",
volume = "891",
pages = "259--295",
year = "2015"
}

@article{Mitev:2012vt,
author = "Mitev, Vladimir and Staudacher, Matthias and Tsuboi, Zengo",
title = "{The Tetrahedral Zamolodchikov Algebra and the ${AdS_5\times S^5}$ S-matrix}",
eprint = "1210.2172",
archivePrefix = "arXiv",
primaryClass = "math-ph",
doi = "10.1007/s00220-017-2905-y",
journal = "Commun. Math. Phys.",
volume = "354",
number = "1",
pages = "1--30",
year = "2017"
}

@article{Hoare:2014kma,
author = "Hoare, Ben and Pittelli, Antonio and Torrielli, Alessandro",
title = "{Integrable S-matrices, massive and massless modes and the $AdS_2 \times S^2$ superstring}",
eprint = "1407.0303",
archivePrefix = "arXiv",
primaryClass = "hep-th",
reportNumber = "HU-EP-14-28, DMUS--MP--14-05",
doi = "10.1007/JHEP11(2014)051",
journal = "JHEP",
volume = "11",
pages = "051",
year = "2014"
}

@article{Metsaev:1998it,
author = "Metsaev, R. R. and Tseytlin, Arkady A.",
title = "{Type IIB superstring action in $AdS_5 \times S^5$ background}",
eprint = "hep-th/9805028",
archivePrefix = "arXiv",
reportNumber = "FIAN-TD-98-21, IMPERIAL-TP-97-98-44, NSF-ITP-98-055",
doi = "10.1016/S0550-3213(98)00570-7",
journal = "Nucl. Phys. B",
volume = "533",
pages = "109--126",
year = "1998"
}

@article{Delduc:2013qra,
author = "Delduc, Francois and Magro, Marc and Vicedo, Benoit",
title = "{An integrable deformation of the $AdS_5 \times S^5$ superstring action}",
eprint = "1309.5850",
archivePrefix = "arXiv",
primaryClass = "hep-th",
doi = "10.1103/PhysRevLett.112.051601",
journal = "Phys. Rev. Lett.",
volume = "112",
number = "5",
pages = "051601",
year = "2014"
}

@article{Klimcik:2002zj,
author = "Klim\v{c}\'{i}k, Ctirad",
title = "{Yang-Baxter sigma models and dS/AdS T duality}",
eprint = "hep-th/0210095",
archivePrefix = "arXiv",
reportNumber = "IML-02-XY",
doi = "10.1088/1126-6708/2002/12/051",
journal = "JHEP",
volume = "12",
pages = "051",
year = "2002"
}

@article{Cherednik:1981df,
author = "Cherednik, I. V.",
title = "{Relativistically Invariant Quasiclassical Limits of Integrable Two-dimensional Quantum Models}",
doi = "10.1007/BF01086395",
journal = "Theor. Math. Phys.",
volume = "47",
pages = "422--425",
year = "1981"
}

@article{Babichenko:2009dk,
author = "Babichenko, A. and Stefanski, Jr., B. and Zarembo, K.",
title = "{Integrability and the AdS(3)/CFT(2) correspondence}",
eprint = "0912.1723",
archivePrefix = "arXiv",
primaryClass = "hep-th",
reportNumber = "ITEP-TH-59-09, LPTENS-09-36, UUITP-25-09",
doi = "10.1007/JHEP03(2010)058",
journal = "JHEP",
volume = "03",
pages = "058",
year = "2010"
}

@article{Wulff:2014kja,
author = "Wulff, Linus",
title = "{Superisometries and integrability of superstrings}",
eprint = "1402.3122",
archivePrefix = "arXiv",
primaryClass = "hep-th",
reportNumber = "IMPERIAL-TP-LW-2014-01",
doi = "10.1007/JHEP05(2014)115",
journal = "JHEP",
volume = "05",
pages = "115",
year = "2014"
}

@article{Klimcik:2008eq,
author = "Klim\v{c}\'{i}k, Ctirad",
title = "{On integrability of the Yang-Baxter sigma-model}",
eprint = "0802.3518",
archivePrefix = "arXiv",
primaryClass = "hep-th",
doi = "10.1063/1.3116242",
journal = "J. Math. Phys.",
volume = "50",
pages = "043508",
year = "2009"
}

@article{Seibold:2019dvf,
author = "Seibold, Fiona K.",
title = "{Two-parameter integrable deformations of the $AdS_3 \times S^3 \times T^4$ superstring}",
eprint = "1907.05430",
archivePrefix = "arXiv",
primaryClass = "hep-th",
doi = "10.1007/JHEP10(2019)049",
journal = "JHEP",
volume = "10",
pages = "049",
year = "2019"
}

@article{Borsato:2023oru,
    author = "Borsato, Riccardo and Driezen, Sibylle and Hoare, Ben and Retore, Ana L. and Seibold, Fiona K.",
    title = "{Inequivalent light-cone gauge-fixings of strings on $AdS_n \times S^n$ backgrounds}",
    eprint = "2312.17056",
    archivePrefix = "arXiv",
    primaryClass = "hep-th",
    month = "12",
    year = "2023"
}

@article{Sundin:2014ema,
author = "Sundin, Per and Wulff, Linus",
title = "{One- and two-loop checks for the $AdS_3 \times S^3 \times T^4$ superstring with mixed flux}",
eprint = "1411.4662",
archivePrefix = "arXiv",
primaryClass = "hep-th",
reportNumber = "IMPERIAL-TP-LW-2014-03",
doi = "10.1088/1751-8113/48/10/105402",
journal = "J. Phys. A",
volume = "48",
number = "10",
pages = "105402",
year = "2015"
}

@article{Borsato:2013qpa,
author = "Borsato, Riccardo and Ohlsson Sax, Olof and Sfondrini, Alessandro and Stefa\'nski, Bogdan and Torrielli, Alessandro",
title = "{The all-loop integrable spin-chain for strings on AdS$_3 \times S^3 \times T^4$: the massive sector}",
eprint = "1303.5995",
archivePrefix = "arXiv",
primaryClass = "hep-th",
reportNumber = "SPIN-13-05, DMUS-MP-13-08, ITP-UU-13-08",
doi = "10.1007/JHEP08(2013)043",
journal = "JHEP",
volume = "08",
pages = "043",
year = "2013"
}

@article{Frolov:2021fmj,
author = "Frolov, Sergey and Sfondrini, Alessandro",
title = "{New dressing factors for AdS3/CFT2}",
eprint = "2112.08896",
archivePrefix = "arXiv",
primaryClass = "hep-th",
doi = "10.1007/JHEP04(2022)162",
journal = "JHEP",
volume = "04",
pages = "162",
year = "2022"
}

@article{Frolov:2021zyc,
author = "Frolov, Sergey and Sfondrini, Alessandro",
title = "{Massless S matrices for AdS3/CFT2}",
eprint = "2112.08895",
archivePrefix = "arXiv",
primaryClass = "hep-th",
doi = "10.1007/JHEP04(2022)067",
journal = "JHEP",
volume = "04",
pages = "067",
year = "2022"
}

@article{Hoare:2013lja,
author = "Hoare, B. and Stepanchuk, A. and Tseytlin, A. A.",
title = "{Giant magnon solution and dispersion relation in string theory in $AdS_3 \times S^3 \times T^4$ with mixed flux}",
eprint = "1311.1794",
archivePrefix = "arXiv",
primaryClass = "hep-th",
reportNumber = "IMPERIAL-TP-AS-2013-01, HU-EP-13-56",
doi = "10.1016/j.nuclphysb.2013.12.011",
journal = "Nucl. Phys. B",
volume = "879",
pages = "318--347",
year = "2014"
}

@article{Hoare:2021dix,
author = "Hoare, Ben",
title = "{Integrable deformations of sigma models}",
eprint = "2109.14284",
archivePrefix = "arXiv",
primaryClass = "hep-th",
doi = "10.1088/1751-8121/ac4a1e",
journal = "J. Phys. A",
volume = "55",
number = "9",
pages = "093001",
year = "2022"
}

@article{OhlssonSax:2023qrk,
author = "Ohlsson Sax, Olof and Riabchenko, Dmitrii and Stefa\'nski, Bogdan",
title = "{Worldsheet kinematics, dressing factors and odd crossing in mixed-flux AdS3 backgrounds}",
eprint = "2312.09288",
archivePrefix = "arXiv",
primaryClass = "hep-th",
reportNumber = "NORDITA 2023-078",
month = "12",
year = "2023"
}

@article{Kawaguchi:2014qwa,
author = "Kawaguchi, Io and Matsumoto, Takuya and Yoshida, Kentaroh",
title = "{Jordanian deformations of the $AdS_5 \times S^5$ superstring}",
eprint = "1401.4855",
archivePrefix = "arXiv",
primaryClass = "hep-th",
reportNumber = "KUNS-2477, ITP-UU-14-05, SPIN-14-05",
doi = "10.1007/JHEP04(2014)153",
journal = "JHEP",
volume = "04",
pages = "153",
year = "2014"
}

@article{Berkovits:1999zq,
author = "Berkovits, N. and Bershadsky, M. and Hauer, T. and Zhukov, S. and Zwiebach, B.",
title = "{Superstring theory on $AdS_2 \times S^2$ as a coset supermanifold}",
eprint = "hep-th/9907200",
archivePrefix = "arXiv",
reportNumber = "IFT-P-060-99, HUTP-99-A044, MIT-CTP-2878, CTP-MIT-2878",
doi = "10.1016/S0550-3213(99)00683-5",
journal = "Nucl. Phys. B",
volume = "567",
pages = "61--86",
year = "2000"
}

@article{Zarembo:2009au,
author = "Zarembo, K.",
title = "{Worldsheet spectrum in AdS(4)/CFT(3) correspondence}",
eprint = "0903.1747",
archivePrefix = "arXiv",
primaryClass = "hep-th",
reportNumber = "ITEP-TH-11-09, LPTENS-09-05, UUITP-08-09",
doi = "10.1088/1126-6708/2009/04/135",
journal = "JHEP",
volume = "04",
pages = "135",
year = "2009"
}

@article{Kruczenski:2004cn,
author = "Kruczenski, Martin and Tseytlin, Arkady A.",
title = "{Semiclassical relativistic strings in S**5 and long coherent operators in N=4 SYM theory}",
eprint = "hep-th/0406189",
archivePrefix = "arXiv",
reportNumber = "BRX-TH-543",
doi = "10.1088/1126-6708/2004/09/038",
journal = "JHEP",
volume = "09",
pages = "038",
year = "2004"
}

@article{Demulder:2023bux,
author = "Demulder, Saskia and Driezen, Sibylle and Knighton, Bob and Oling, Gerben and Retore, Ana L. and Seibold, Fiona K. and Sfondrini, Alessandro and Yan, Ziqi",
title = "{Exact approaches on the string worldsheet}",
eprint = "2312.12930",
archivePrefix = "arXiv",
primaryClass = "hep-th",
reportNumber = "NORDITA 2023-083",
month = "12",
year = "2023"
}

@article{Frolov:2019nrr,
author = "Frolov, Sergey",
title = "{$T\overline{T}$ Deformation and the Light-Cone Gauge}",
eprint = "1905.07946",
archivePrefix = "arXiv",
primaryClass = "hep-th",
reportNumber = "TCD-MATH-19-06",
doi = "10.1134/S0081543820030098",
journal = "Proc. Steklov Inst. Math.",
volume = "309",
pages = "107--126",
year = "2020"
}

@article{Baggio:2018gct,
author = "Baggio, Marco and Sfondrini, Alessandro",
title = "{Strings on NS-NS Backgrounds as Integrable Deformations}",
eprint = "1804.01998",
archivePrefix = "arXiv",
primaryClass = "hep-th",
doi = "10.1103/PhysRevD.98.021902",
journal = "Phys. Rev. D",
volume = "98",
number = "2",
pages = "021902",
year = "2018"
}

\end{bibtex}

\bibliographystyle{nb}
\bibliography{\jobname}

\end{document}